\documentclass[notitlepage,a4paper,aps,prd,onecolumn,superscriptaddress,nofootinbib,groupedaddress]{revtex4}
\usepackage{enumerate}
\usepackage{amsmath}
\usepackage{amsfonts}
\usepackage{amssymb}
\usepackage[utf8]{inputenc}
\usepackage[T1]{fontenc}
\usepackage{mathtools}
\usepackage{wasysym}
\usepackage{accents,float}
\usepackage{ulem}

\newcommand{\lc}[1]{\accentset{\circ}{#1}}
\newcounter{subeq}

\DeclareMathAlphabet{\mathds}{U}{BOONDOX-ds}{m}{n}
\usepackage[dvipsnames]{xcolor}

\usepackage[colorlinks=true]{hyperref}
\usepackage{amsthm}

\theoremstyle{definition}

\theoremstyle{plain}

\newcommand{\udt}[3]{#1^{#2}_{\phantom{#2}#3}}

\newcommand{\dut}[3]{#1_{#2}^{\phantom{#2}#3}}

\allowdisplaybreaks

\begin{document}
	\title{Exploring Axial Symmetry in Modified Teleparallel Gravity}
	
	\author{Sebastian Bahamonde}
	\email{sbahamonde@ut.ee}
	\affiliation{Laboratory of Theoretical Physics, Institute of Physics, University of Tartu, W. Ostwaldi 1, 50411 Tartu, Estonia.}
	\affiliation{Laboratory for Theoretical Cosmology, Tomsk State University of
		Control Systems and Radioelectronics, 634050 Tomsk, Russia (TUSUR)}
	
	\author{Jorge Gigante Valcarcel}
	\email{jorge.gigante.valcarcel@ut.ee}
	\affiliation{Laboratory of Theoretical Physics, Institute of Physics, University of Tartu, W. Ostwaldi 1, 50411 Tartu, Estonia.}
	
	\author{Laur J\"arv}
	\email{laur.jarv@ut.ee}
	\affiliation{Laboratory of Theoretical Physics, Institute of Physics, University of Tartu, W. Ostwaldi 1, 50411 Tartu, Estonia.}

	\author{Christian Pfeifer}
	\email{christian.pfeifer@ut.ee}
	\affiliation{Laboratory of Theoretical Physics, Institute of Physics, University of Tartu, W. Ostwaldi 1, 50411 Tartu, Estonia.}
	
	\begin{abstract}
		Axially symmetric spacetimes play an important role in the relativistic description of rotating astrophysical objects like black holes, stars, etc. In gravitational theories that venture beyond the usual Riemannian geometry by allowing independent connection components, the notion of symmetry concerns, not just the metric, but also the connection. As discovered recently, in teleparallel geometries, axial symmetry can be realised in two branches, while only one of these has a continuous spherically symmetric limit. In the current paper, we consider a very generic $f(T,B,\phi,X)$ family of teleparallel gravities, whose action depends on the torsion scalar $T$ and the boundary term $B$, as well as a scalar field $\phi$ with its kinetic term $X$. As the field equations can be decomposed into symmetric and antisymmetric (spin connection) parts, we thoroughly analyse the antisymmetric equations and look for solutions of axial spacetimes which could be used as ans\"atze to tackle the symmetric part of the field equations. In particular, we find solutions corresponding to a generalisation of the Taub-NUT metric, and the slowly rotating Kerr spacetime. Since this work also concerns a wider issue of how to determine the spin connection in teleparallel gravity, we also show that the method of ``turning off gravity'' proposed in the literature, does not always produce a solution to the antisymmetric equations.
	\end{abstract}
	
	\maketitle
	
	\section{Introduction}\label{sec:intro}
	It took hardly a month since the publication of Einstein's theory of general relativity (GR) for Karl Schwarzschild to produce a solution of the field equations in the spherically symmetric case. However, many interesting astrophysical objects from stars and planets to black holes exhibit some rotation, i.e.\ possess just stationary axial symmetry and cannot be described by a spherical spacetime precisely. In general relativity, it took almost six decades until Ezra T. Newman and Roy Kerr worked out rotating solutions \cite{Teukolsky:2014vca}, and then a bit more than a dozen years to map out the full Pleba\'nski-Demia\'nski family of axially symmetric spacetimes \cite{Griffiths:2009dfa,Stephani:2003tm}. In $f(\lc{R})$ and other extensions of general relativity the known exact axial solutions are few and far between (e.g.\ \cite{Cembranos:2011sr,Filippini:2017kov,Chauvineau:2018zjy,Ding:2019mal,Jusufi:2019caq,Guerrero:2020azx,BenAchour:2020fgy,Anson:2020trg}), mostly recovered in the slow rotation limit \cite{Konno:2009kg, Yunes:2009hc, Pani:2009wy, Pani:2011gy, Ayzenberg:2014aka,Maselli:2015tta,Cano:2019ore}, or approached by the continued fraction expansion \cite{Konoplya:2016jvv,Konoplya:2018arm}. However, conceptually the procedure for finding the solutions is clear. The symmetry is encoded in the Killing vectors which leads to an ansatz for the metric, and free functions in the ansatz can then be fixed by the field equations.
	
	Teleparallel gravity uses a geometric identity
	whereby the Levi-Civita Ricci scalar $\lc{R}$ of the Einstein-Hilbert action can be rewritten in terms of the torsion scalar $T$ and a total divergence of the torsion tensor. The latter constitutes a boundary term $B$, which does not affect the equations of motion. The action given by the torsion scalar is called teleparallel equivalent of general relativity (TEGR), as adopting the so-called Weitzenböck connection of vanishing curvature (and vanishing nonmetricity) grants distant parallel transport of vectors \cite{Aldrovandi:2013wha,Weitzenbock1923}. TEGR can be extended to $f(T)$ \cite{Ferraro:2006jd,Bengochea:2008gz,Linder:2010py,Hohmann:2017duq} and further $f(T, B, \phi, X)$ modifications, where $\phi$ is a scalar field and $X$ its kinetic term \cite{Geng:2011aj,Bahamonde:2015zma,Bahamonde:2015hza,Hohmann:2018rwf,Hohmann:2018dqh,Abedi:2018lkr,Bahamonde:2019shr}. Interestingly, this includes a very broad class of theories, among others the $f(\mathring{R},\phi,X)$
	extensions of standard general relativity. In this picture, the properties of gravity can be attributed to torsion instead of curvature. The price to pay is the introduction of additional connection components, extra to the usual Levi-Civita ones which follow from the metric.

	Thus, in contrast to general relativity, teleparallel gravities face the  problem of how to determine the extra connection. By definition, the connection must (i) be flat (giving vanishing curvature), and obviously also (ii) solve the field equations arising from the variation of the action with respect to the flat connection. Incidentally, the connection equations coincide with the equations for the antisymmetric tetrad components \cite{Golovnev:2017dox,Hohmann:2017duq,Hohmann:2018rwf}, but are identically satisfied in TEGR \cite{Aldrovandi:2013wha}. Arguably, the connection should also better (iii) obey the same symmetry as the metric \cite{Hohmann:2019nat,Coley:2019zld}. Here one should realise that it is conceivable to consider a configuration where the connection or equivalently the torsion tensor possesses less or different symmetry than the metric, although the physical relevance of such situations is not clear. Further on, it makes also sense to prefer such a connection which enables to (iv) define meaningful conserved charges in the asymptotics (like mass or angular momentum) \cite{Lucas:2009nq,Obukhov:2006sk,Krssak:2015rqa,Emtsova:2019moq}. Another idea to fix the connection is that it should (v) renormalise the action in the IR \cite{Krssak:2015rqa,Krssak:2015lba}, while related proposals to determine the connection are to require it to vanish in the limit when ``gravity is turned off'' \cite{Krssak:2015rqa,Krssak:2018ywd}. 
	
	The flatness condition is easy to solve by employing the tetrad formalism, whereby one can assume the so-called Weitzenb\"ock gauge, where vanishing spin connection immediately implies vanishing curvature. Yet, in that case, one has still to figure out which Lorentz frame belongs to the vanishing spin connection, or, the other way around, which one is the correct tetrad to which one associates the vanishing spin connection.
	Entertaining the terminology of Ref.\ \cite{Tamanini:2012hg} we may call a tetrad ``good'' if it solves the antisymmetric field equations with vanishing spin connection. Indeed, looking for a ``good'' tetrad is quite often a useful approach in trying to solve the equations. If a ``good'' tetrad is found, then applying local Lorentz transformations will not just transform the tetrad but typically also introduce non-vanishing spin connection in a covariant manner \cite{Krssak:2015oua,Krssak:2018ywd}. One should keep in mind that any tetrad and spin connection pair related to the ``good'' tetrad by a local Lorentz transformation will solve the antisymmetric field equations. 
	
	For the spherical symmetry in $f(T,B,\phi,X)$ gravity the ``good'' tetrad is known \cite{Ferraro:2011ks}. It satisfies all the points above, i.e.\ by definition, it is associated to a vanishing flat spin connection and solves the antisymmetric equations \cite{Tamanini:2012hg}, but also obeys the symmetry \cite{Hohmann:2019nat}, defines the correct mass in the asymptotics \cite{Emtsova:2019moq}, and renormalises the IR action \cite{Krssak:2015rqa}. However, as far as rotating solutions and axial symmetry are concerned, the literature remains lacking a satisfactory result. The early tetrad expressions of the Kerr metric \cite{Lucas:2009nq, Krssak:2015rqa} aimed to give correct mass and to renormalise the action at IR, do not solve the connection field equations and hence can at best pertain to TEGR only. 
	The other tetrad for Kerr spacetime found by Bejarano \textit{et al.}\ in the null tetrad formalism \cite{Bejarano:2014bca} does solve the field equations trivially since it has vanishing $T$ and $B$. However, as we argue in this paper, it has a subtle issue with symmetry. Namely, owing to group theory considerations, teleparallel connections with axial symmetry come in two branches \cite{Hohmann:2019nat}. Only the first, regular branch can be continuously related to the spherically symmetric case mentioned above \cite{Ferraro:2011ks,Tamanini:2012hg}, while the connections in the other branch (including the solution in \cite{Bejarano:2014bca}) fail to exhibit spherical symmetry in the limit where the corresponding metric becomes spherical. There is also a solution found by some of the present authors earlier \cite{Jarv:2019ctf}, which satisfies the antisymmetric field equations and belongs to the regular branch of axial symmetry, but is rather limited in the sense that it does not incorporate the possibility of the Kerr metric. In the literature, one may come across a few other proposals for rotating solutions in teleparallel gravities, however, these fall short of fulfilling the other conditions except flatness.
	
	In the present work, we give an account of an effort to describe rotating geometries in teleparallel gravities. The broader aim is twofold. The first task would be to determine the teleparallel connection components that can go together with the Kerr metric and obey the conditions (i) - (v) above. The second aim is to get hold of an ansatz for a rotating ``good'' tetrad, i.e.\ a tetrad in Weitzenb\"ock gauge that obeys the symmetry and solves the antisymmetric equations independent of the function $f$. This ansatz could then be substituted into the symmetric equations to find solutions in different theories belonging to the $f(T, B, \phi, X)$ family. Both aims remain yet to be reached in full glory, but nevertheless, the current paper is able to report on several interesting results and provide the groundwork for further investigations. After recalling a few key formulae of teleparallel gravity in Sec.~\ref{ssec:TPGrav}, we explain how axial symmetry can be realised by Weitzenb\"ock tetrads (i.e.\ tetrads associated with vanishing spin connection) belonging to two branches in Sec.~\ref{ssec:2}. Then in Sec.~\ref{ssec:reular_branch} we consider vacuum Pleba\'nski-Demia\'nski geometry and propose a generic form for a ``good'' tetrad which pertains to the regular branch and automatically satisfies all antisymmetric field equations except one. There are different ways how to tackle the remaining equation, and by treating it case by case we are able to derive different solutions such as e.g.\ a generalisation of the solution of Ref.~\cite{Jarv:2019ctf}; a solution that accommodates Taub-NUT spacetime; a solution that corresponds to the Kerr metric in the slow rotation expansion. Afterward, in Sec.~\ref{sec:2ndbranch} we propose a generic form for a ``good'' tetrad which pertains to the other branch but will not consider it further since it falls short of the spherical symmetry limit. In this section, we also comment on the time-dependent Kerr tetrad found in Ref.~\cite{Bejarano:2014bca}. Lastly, in Sec.~\ref{ssec:determinig_connection} we show how the method outlined in Ref.~\cite{Emtsova:2019moq} to obtain a teleparallel connection from a metric produces a ``good'' tetrad in the Taub-NUT case, but not in the Kerr or C-metric case. Sec.~\ref{sec:conclusion} offers a final discussion. The appendices \ref{ssec:torsion_scalar} and \ref{ssec:kerr_perturbations} list some long but necessary expressions for the axially symmetric torsion scalar and boundary term.
	
	As a remark for readers who are familiar with metric-affine gravity, it may be mentioned that teleparallel framework has some similarities, but also differences. In both contexts the notion of symmetry encompasses both the metric and independent connection \cite{Rauch:1981tva,Hohmann:2019fvf,Bahamonde:2020fnq,Hehl:1994ue}. However, the rotating solutions found in metric-affine gravities, e.g.\ \cite{Bakler:1988nq,Hehl:1999sb,Baekler:2006de}, will not likely reduce to meaningful teleparallel configurations, since the curvature tensor generally plays a dynamical role, while in the teleparallel case the connection is necessarily flat. Hence when setting curvature to zero for an arbitrary metric-affine solution its key features will be lost. 
	
	Throughout the paper, we denote $h^{A}{}_{\mu}$ and $h_{A}{}^{\mu}$ for the tetrad and its inverse, respectively, where capital Latin indices refer to tangent space indices and Greek to spacetime indices. Both indices run from 0,..,3.  In addition, over-circles $\circ$ on top denotes quantities computed with the Levi-Civita connection. Quantities without any symbol on top denote that they are computed with the 
	Weitzenb\"ock connection (teleparallel). 
	
	Our signature convention is $(+,-,-,-)$, $\eta$ denotes the Minkowski metric with components $\eta_{AB} = \textrm{diag}(+,-,-,-)$ and we work in units where $G=c=1$.
	
	\section{Teleparallel theories of gravity}\label{ssec:TPGrav}
	General relativity is constructed from the unique torsionless connection satisfying the metric compatibility condition, which is known as the Levi-Civita connection defined by the Christoffel symbols $\lc{\Gamma}^{\alpha}{}_{\mu\nu} = \frac{1}{2}g^{\alpha\rho}(\partial_\mu g_{\rho\nu}+\partial_\nu g_{\rho\mu} - \partial_\rho g_{\mu\nu})$. On the other hand, torsional teleparallel gravity assumes a specific connection known as the Weitzenb\"ock connection $\Gamma^{\alpha}{}_{\mu\nu}$ which is torsionful ($T^{\alpha}{}_{\mu\nu}\neq0$), metric compatible ($\nabla_\alpha g_{\mu\nu}=0$) and curvatureless ($R^{\alpha}{}_{\mu\nu\beta}=0$)~\cite{Aldrovandi:2013wha,Weitzenbock1923}. In this framework, the fundamental dynamical objects are tuples $(h^A{}_\mu,\omega^A{}_{B\mu})$ consisting of a tetrad $h^A{}_\mu$, acting as soldering agents from the spacetime manifold (Greek indices) and the tangent space (capital Latin indices), and a spin connection  $\omega^A{}_{B\mu}$ which can be seen as a pure gauge quantity. The metric and its inverse can be reconstructed from the tetrad fields using the following relationships,
	\begin{align}
	g_{\mu\nu} &= \udt{h}{A}{\mu}\udt{h}{B}{\nu} \eta_{AB}\,,\quad
	\eta_{AB} = \dut{h}{A}{\mu}\dut{h}{B}{\nu} g_{\mu\nu}\,,
	\end{align}
	where $\eta_{AB}$ is the Minkowski metric and $\dut{h}{A}{\mu}$ is the inverse of the tetrad satisfying $\dut{h}{A}{\mu}\udt{h}{A}{\nu}=\delta^\mu_\nu$. 
	The torsion tensor is then defined as the antisymmetric part of the Weitzenb\"ock connection:
	\begin{align}
	T^A{}_{\mu\nu} =2\,\Gamma^{A}{}_{[\nu\mu]}= \partial_{\mu}h^A{}_{\nu}-\partial_{\nu}h^A{}_{\mu} + \omega^A{}_{B\mu} h^B{}_{\nu}-\omega^A{}_{B\nu} h^B{}_{\mu}\,,
	\end{align}
	which is covariant under local Lorentz transformation and the spin connection is given by~\cite{Krssak:2015oua,Golovnev:2017dox}
	\begin{equation}
	\omega^A{}_{B\mu}=\Lambda^A{}_C\, \partial_\mu (\Lambda^{-1})^C{}_{B}\,,
	\label{eq: local Lorentz transformation}
	\end{equation}
	where $\Lambda^A{}_B$ is the Lorentz matrix. It means that the above quantity is a pure gauge object. This can be seen after taking local Lorentz transformations for both the tetrads and the spin connection which yields in
	\begin{equation}
	h'{}^A{}{\mu}=\Lambda'^A{}_{B}h^B_{}{ \mu}, \quad \quad
	\omega'{}^A{}_{B\mu}=\Lambda'^A{}_{C}\,\omega^C{}_{D\mu}(\Lambda'^{-1})^{D}{}_B+\Lambda'^A{}_{C}\, \partial_\mu (\Lambda'^{-1})^{C}{}_B = \tilde \Lambda^A{}_C \partial_{\mu} (\tilde \Lambda^{-1})^{C}{}_B,\label{lortrans}
	\end{equation}
	with $\tilde \Lambda^A{}_C = \Lambda'^A{}_B \Lambda^B{}_C$. Thus, in all frames, the spin connection remains flat and fully determined by a Lorentz matrix. Hence, any teleparallel theory has the tetrads and spin connection as their basic variables $(h^A{}_\mu,\omega^A{}_{B\mu})$, but the latter one can be always gauged away by choosing a specific frame, taking $\Lambda' = \Lambda^{-1}$ in \eqref{lortrans}, where the spin connection coefficients $\omega^A{}_{B\mu}$ vanish. The tetrad belonging to the tuple $(h^A{}_\mu,0)$, i.e.\ the tetrad with vanishing spin connection, is called a Weitzenb\"ock tetrad. In this so-called Weitzenb\"ock gauge the torsion tensor just becomes
	\begin{align}
	T^A{}_{\mu\nu} =2 \partial_{[\mu}h^A{}_{\nu]}\,.
	\end{align}
	
	One of the most interesting aspects of teleparallel gravity is that it is possible to construct a theory which is equivalent to GR, by considering the following action
	\begin{equation}
	\mathcal{S}_{\rm TEGR}=\int d^4x\, h\, \Big[\frac{1}{2\kappa^2}T+L_{\rm m}\Big]\,,\label{action0}
	\end{equation}
	where $L_{\rm m}$ is the matter Lagrangian, $h=\textrm{det}(h^{A}{}_\mu)=\sqrt{-g}$, $\kappa^2=8\pi G$ and $T$ is known as the torsion scalar which is a specific combination of contractions of the torsion tensor, namely
	\begin{equation}
	T=\frac{1}{2}S^{\alpha\mu\nu}T_{\alpha \mu \nu}=\frac{1}{4}T^{\mu\nu\rho}T_{\mu\nu\rho} + \frac{1}{2}T^{\mu\nu\rho}T_{\rho\nu\mu} - T_{\rho}T^{\rho}\,,\label{defT}
	\end{equation}
	where $T^\mu{}_{\mu\rho}=T_\rho$ and we have also defined the superpotential as \begin{eqnarray}
	S_{\rho}{}^{\mu\nu} &=&K^{\mu\nu}{}_{\rho}-\delta_{\rho}^{\mu}T_{\sigma}{}^{\sigma\nu}+\delta_{\rho}^{\nu}T_{\sigma}{}^{\sigma\mu}=-S_{\rho}{}^{\nu\mu}\,,
	\end{eqnarray}
	and the contortion tensor as 
	\begin{eqnarray}
	K^{\rho}{}_{\mu\nu} & =&\Gamma^{\rho}{}_{\mu\nu}-\lc{\Gamma}^{\rho}{}_{\mu\nu}=\frac{1}{2}\left(T_{\mu}{}^{\rho}{}_{\nu}+T_{\nu}{}^{\rho}{}_{\mu}-T^{\rho}{}_{\mu\nu}\right)\,.
	\end{eqnarray}
	
	After imposing that the curvature tensor is zero $R^\alpha{}_{\beta\mu\nu}=0$, one can show that the torsion scalar $T$ and the Ricci scalar $\lc{R}$ differ from each other by a boundary term $B$:
	\begin{equation}\label{TEGR_L}
	R=\lc{R} +T-\frac{2}{h}\partial_{\mu}\left(hT^{\sigma}{}_{\sigma}{}^{\mu}\right)=0 \quad \Rightarrow \quad \lc{R} = -T + \frac{2}{h}\partial_{\mu}\left(hT^{\sigma}{}_{\sigma}{}^{\mu}\right) := -T + B\,.
	\end{equation}
	This means that after taking variations with respect to the tetrads, the corresponding symmetric field equations coming from the action~\eqref{action0} are identical to the Einstein field equations while the antisymmetric equations are identically satisfied. For this reason, this theory is known as ``Teleparallel equivalent of General Relativity'' (TEGR). 
	
	Since the action~\eqref{action0} gives us the same dynamics as GR, one can then modify it in different ways to construct modified teleparallel theories of gravity. The most straightforward modification is $f(T)$ gravity where one replaces $T$ in the action to an arbitrary function which depends on the torsion scalar~\cite{Cai:2015emx,Krssak:2018ywd,Ferraro:2006jd}. From~\eqref{TEGR_L} one directly notices that $f(T)$ is not equivalent to $f(\lc{R})$ gravity which is the generalisation of the Einstein-Hilbert action from $\lc{R}$ to an arbitrary function $f(\lc{R})$. Moreover, $f(T)$ gravity is a second order theory whereas $f(\lc{R})$ is a fourth other theory. 
	One can then extend $f(T)$ gravity by also adding the boundary term $B$ in the action, which leads to $f(T,B)$ gravity~\cite{Bahamonde:2015zma}. This theory contains both $f(T)$ and $f(\lc{R})$ by taking the limits $f(T,B)=f(T)$ and $f(T,B)=f(-T+B)$, respectively. The field equations of this theory are fourth order as in $f(\lc{R})$ gravity.
	
	In order to encapsulate different modified teleparallel theories of gravity, we will then consider a generalisation of the theories described before by adding a scalar field $\phi$, namely,
	\begin{align}
	\mathcal{S}_{ f(T,B,\phi,X)} = \int 
	d^4x\, h\, \left[ 
	\frac{1}{2\kappa^2}f(T,B,\phi,X) + L_{\rm m}
	\right] \,,\label{action}
	\end{align}
	where now the function $f$ also depends on a scalar field $\phi$ and its
	kinetic term $X\equiv -(\epsilon/2) \, g^{\mu\nu}\partial_{\mu}\phi\partial_{\nu}\phi=-(\epsilon/2)(\lc{\nabla}\phi)^2$, so that, if $\epsilon=1$ ($\epsilon=-1$) we have a canonical(phantom) scalar field. This action represents a very rich range of modified theories of gravity. Indeed, some of these theories are dynamically equivalent to non-teleparallel theories. For example, if we choose $f(T,B,\phi,X)=f(-T+B,\phi,X)=f(\lc{R},\phi,X)$, we recover the action studied in \cite{Beltran:2015hja} in a cosmological framework which is a generalisation of curvature-based models with a scalar field (see~\cite{Nojiri:2017ncd} for a review about them). Furthermore, several other teleparallel scalar-tensor type theories are part of this action such as teleparallel dark energy~\cite{Geng:2011aj,Xu:2012jf}, theories with couplings between the boundary term and the scalar field~\cite{Zubair:2016uhx,Bahamonde:2015hza,Bahamonde:2016jqq}, tachyonic models~\cite{Bahamonde:2019gjk} or more general scalar-tensor type theories such as some of the ones described in~\cite{Abedi:2018lkr,Hohmann:2018ijr,Hohmann:2018vle,Hohmann:2018dqh,Hohmann:2018rwf}.
	
	In general, the field equations to the action \eqref{action} are obtained by variation with respect to the tetrad components as well as by variation with respect to the flat spin connection components. It then turns out that, for general teleparallel theories of gravity, the antisymmetric part of the tetrad field equation is equivalent to the spin connection field equation~\cite{Golovnev:2017dox,Hohmann:2017duq,Hohmann:2018rwf}. This shows one more time that the spin connection is a pure gauge quantity and it suffices to derive the tetrad field equations. For \eqref{action} the tetrad field equations in Weitzenb\"ock gauge are~\cite{Bahamonde:2015zma}
	\begin{multline}
	2\delta_{\nu}^{\lambda}\lc{\Box} f_{B}-2\lc{\nabla}^{\lambda}\lc{\nabla}_{\nu}f_{B}+
	B f_{B}\delta_{\nu}^{\lambda} + 
	4\Big[(\partial_{\mu}f_{B})+(\partial_{\mu}f_{T})\Big]S_{\nu}{}^{\mu\lambda}
	\\
	+4h^{-1}h^{A}{}_{\nu}\partial_{\mu}(h S_{A}{}^{\mu\lambda})f_{T} - 
	4 f_{T}T^{\sigma}{}_{\mu \nu}S_{\sigma}{}^{\lambda\mu} - 
	f \delta_{\nu}^{\lambda}+ \epsilon\, f_{X}\partial^{\lambda}\phi \partial_{\nu}\phi =2\kappa^2 \mathcal{T}_{\nu}^{\lambda}\,,
	\label{fieldeq}
	\end{multline}
	where $\mathcal{T}_{\nu}^{\lambda}$ is the standard energy momentum tensor that it was defined from the Lagrangian matter as
	\begin{align}
	\mathcal{T}_{\mu\nu} & :=\frac{-2}{\sqrt{-g}}\frac{\delta (h\, L_{m})}{\delta g^{\mu\nu}}=h^{A}{}_{\mu}\left(\frac{1}{h}\frac{\delta (h\, L_{m})}{\delta h^{A}{}_{\alpha}}\right)g_{\nu\alpha}:=h^{A}{}_{\mu}\mathcal{T}_{A}{}^{\alpha}g_{\nu\alpha}\,,
	\end{align}
	while variations with respect to the scalar field $\phi$ yields,
	\begin{eqnarray}
	\epsilon\,\partial_{\mu}\Big(hf_{X}g^{\mu\nu}\partial_{\nu}\phi\Big)+h f_{\phi}=0\,.\label{fieldeq2}
	\end{eqnarray} 
	Here, $f_X=\partial f/\partial X$, $f_{\phi}=\partial f/\partial \phi$, $f_T=\partial f/\partial T$ and $f_B=\partial f/\partial B$. The antisymmetric part of the field equation~\eqref{fieldeq} becomes 
	\begin{equation}
	E_{[\mu\nu]}:= 4\Big[(\partial_{\rho}f_{B})+(\partial_{\rho}f_{T})\Big]S_{[\mu}{}^{\rho}{}_{\nu]}=\frac{3}{2}T^{\rho}{}_{[\mu\nu}\partial_{\rho]}(f_T+f_B)\,.\label{antitotal}
	\end{equation}
	Let us emphasise again here that the above equation coincides with the variations with respect to the spin connection. One can also perform a local Lorentz transformation to consider the equations, not in Weitzenb\"ock gauge but with the spin connection. This would affect the field equations in the following way: the appearing torsion tensor needs to be expanded including the spin connection generated by the Lorentz transformation and the term $h^{-1} h^A{}_\nu \partial _\mu(hS_A{}^{\mu\nu})$ would generate a spin connection counter term which can be combined with the partial derivative into a covariant Fock-Ivanenko derivative, i.e. $h^A{}_\nu D_\mu(S_A{}^{\mu\nu})$ (see \cite{Aldrovandi:2013wha} for a definition of the Fock-Ivanenko derivative). Having a solution $(h^A{}_\mu,\omega^{A}{}_{B\mu}=0)$, we can obtain solutions in other frames by making a local Lorentz transformation $\Lambda^A{}_B$ (see Eq.~\eqref{lortrans}) and obtain another tuple $(h'^A{}_\mu,\omega'^{A}{}_{B\mu})$.
	
	Note first that if one finds a tetrad for which $T,B$, and $\phi$ are constant, independently of $f$, then the field equations reduce to the TEGR (GR) field equations plus a cosmological constant. Second, if the function $f$ satisfies $f_T = -f_B$, then the theory is dynamically equivalent to $f(\mathring{R},\phi,X)$ gravity.
	
	The main aim is to solve the above antisymmetric equation $ E_{[\mu\nu]}=0$ without choosing the trivial cases where one recovers $f(\lc{R},\phi,X)$ or TEGR. To match the terminology sometimes used in teleparallel gravity, we will label as ``good tetrads''~\cite{Tamanini:2012hg} to those tetrads which solve the antisymmetric field equations~\eqref{antitotal} in the Weitzenb\"ock gauge.
	
	\section{Axial Symmetry in teleparallel gravity}\label{ssec:2}
	In this section let us briefly recall the results on axially symmetric teleparallel geometries from the literature. In particular, we highlight that there exist two branches of axially symmetric Weitzenb\"ock tetrads, of which only the first branch is compatible with a limit to a spherically symmetric teleparallel geometry.
	
	\subsection{The notion of symmetry}
	The class of spacetime symmetries provided by the action of a Lie group on a differentiable manifold in the framework of teleparallelism is based on the invariance of the underlying Cartan geometry modeled by principal bundle automorphisms \cite{Hohmann:2015pva}. In particular, infinitesimal symmetries can be described by the invariance of the geometric structure of the manifold under the flow of a set of $\zeta=1,..,m$ vector fields $Z_\zeta$, which involves the vanishing of the Lie derivative in the direction of $Z_\zeta$ not only on the metric tensor $g$ but also on the affine teleparallel connection coefficients $ \Gamma$:
	\begin{align}\label{eq:symm}
	\mathcal{L}_{Z_\zeta}g_{\mu\nu}=0\,,\quad
	\mathcal{L}_{Z_\zeta}\Gamma^{\lambda}\,_{\mu\nu}=0\,.
	\end{align}
	Expanding the teleparallel affine connection coefficients $\Gamma$ into Levi-Civita and contortion part, the first condition implies that the Levi-Civita part, represented by the Christoffel symbols $\lc{\Gamma}$ of the metric, is straightforwardly preserved by a group of isometries, in virtue of the vanishing of its Lie derivative \cite{Yano1972notes}:
	\begin{equation}
	\mathcal{L}_{Z_\zeta}\lc{\Gamma}^{\lambda}\,_{\mu\nu}=\frac{1}{2}g^{\lambda\rho}\left(\lc{\nabla}_{\mu}\mathcal{L}_{Z_\zeta}g_{\rho\nu}+\lc{\nabla}_{\nu}\mathcal{L}_{Z_\zeta}g_{\rho\mu}-\lc{\nabla}_{\rho}\mathcal{L}_{Z_\zeta}g_{\mu\nu}\right)\,.
	\end{equation}
	This means that the introduction of post-metric degrees of freedom into the affine connection requires its subsequent independent symmetry condition.
	
	For a teleparallel connection, that posses no curvature and is metric compatible, equivalently, the existence of a Lie algebra homomorphism $\lambda$, associated with a global Lie group homomorphism $\Lambda$, which maps the symmetry group into the Lorentz group, allows the mentioned symmetry conditions to be expressed in terms of the tetrad field and the spin connection as follows \cite{Hohmann:2019nat}:
	\begin{align}\label{eq:tpsymm}
	\mathcal{L}_{Z_\zeta}h^{A}\,_{\mu}=-\,\lambda^{A}_{\zeta}{}_{B}h^{B}\,_{\mu} \,,\quad 
	\mathcal{L}_{Z_\zeta}\omega^{A}\,_{B\mu}=\partial_{\mu}\lambda^{A}_{\zeta}{}_{B}+\omega^{A}\,_{C\mu}\lambda^{C}_{\zeta}{}_{B}-\omega^{C}\,_{B\mu}\lambda^{A}_{\zeta}{}_{C}\,.
	\end{align}
	Solving these equations in Weitzenb\"ock gauge, i.e.\ with vanishing spin connection, implies that the Lie algebra homomorphism $\lambda$ cannot depend on spacetime points, $\partial_{\mu}\lambda^{A}_{\zeta}{}_{B}=0$. Hence, the only remaining equation which needs to be solved is the symmetry condition for the tetrad for a fixed choice of $\lambda$. 
	
	Both versions of the symmetry condition, \eqref{eq:symm} or \eqref{eq:tpsymm}, imply for the torsion tensor, expressed through the teleparallel affine connection corresponding to the spin connection, that
	\begin{align}\label{eq:torsym}
	\mathcal{L}_{Z_\zeta}T^\sigma{}_{\mu\nu} = \mathcal{L}_{Z_\zeta} \Gamma^\sigma{}_{[\mu\nu]} = 0\,.
	\end{align}
	Hence, for a teleparallel geometry, the symmetry of tetrad and spin connection implies the symmetry of the torsion tensor.
	
	Depending on which symmetry group is considered, there may exist more than one homomorphism $\lambda$. Different choices of this mapping then lead to different branches of symmetric teleparallel geometries.
	
	In addition, since we are studying a theory with a scalar field, we must also impose that the symmetry conditions are respected by the scalar field $\phi$, which means it must satisfy
	\begin{align}
	\mathcal{L}_{Z_\zeta}\phi = Z_\zeta(\phi) = 0\,.
	\end{align}
	A consequence of this condition is that the kinetic term $X$ is axially symmetric as well.

	\subsection{Two branches of Weitzenb\"ock tetrads}\label{sec:twobranches}
	For the specific case of axial symmetry, the Killing vector $\partial_{\varphi}$ defines a regular two-dimensional timelike surface of fixed points where it vanishes and the Cartan geometry is invariant under the action of the underlying rotation group $SO(2)$ \cite{Stephani:2003tm}. 
	
	For the scalar field the symmetry condition is easily solved by $\phi=\phi(r,\vartheta)$ and thus $X=X(r,\vartheta)$.
	
	For the tetrad, we recall that there exist two different group homomorphisms $\Lambda$, and hence two different Lie algebra homomorphisms $\lambda$, which map the symmetry group into the Lorentz group, leading to two branches of axially symmetric tetrads with vanishing spin connection, all details can be found in Ref.\ \cite{Hohmann:2019nat}.
	
	\subsubsection{Regular branch}\label{ssec:regbranch}
	The nontrivial group homomorphism leads to a non trivial Lie algebra homomorphism $\lambda$ 
	\begin{align}
	\lambda(\partial_\varphi) = 
	\left(
	\begin{array}{cccc}
	0 & 0 & 0 & 0\\
	0 & 0 & -1 & 0\\
	0 & 1 & 0 & 0\\
	0 & 0 & 0 & 0\\
	\end{array}
	\right)
	\end{align}
	and solving \eqref{eq:tpsymm},
	yields the following Weitzenb\"ock gauge tetrad \cite{Hohmann:2019nat}
	\begin{equation}
	h^A{}_\mu=    \left(
	\begin{array}{cccc}
	H_{00} & H_{01} & -H_{02} & H_{03} \\
	H_{10}\cos\varphi-H_{20}\sin\varphi  &H_{11}\cos\varphi - H_{21}\sin\varphi & H_{22}\sin\varphi +H_{12}\cos\varphi  &H_{13}\cos\varphi -H_{23}\sin\varphi  \\
	H_{10}\sin\varphi +H_{20}\cos\varphi  & H_{11}\sin\varphi +H_{21} \cos\varphi & H_{12}\sin\varphi -H_{22}\cos\varphi  & H_{13}\sin\varphi +H_{23}\cos\varphi  \\
	H_{30} & H_{31} & -H_{32} & H_{33} \\
	\end{array}
	\right)\,,
	\label{eq: general axial tetrad}
	\end{equation}
	where $(t,r,\vartheta,\varphi)$ denote spherical coordinates and $\{H_{ij}\}_{i,j=0}^{3}$ are sixteen arbitrary functions depending on $t,r$ and $\vartheta$. Assuming stationarity, that we will do to analyse this branch (see Sec.~\ref{ssec:reular_branch}), the functions will depend only on $r$ and $\vartheta$.
	
	It is straightforward to note that the above tetrad maintains the same structure in Boyer-Lindquist coordinates $(\tilde{t},\tilde{r},\tilde{\vartheta},\tilde{\varphi})$:
	\begin{eqnarray}
	t=\tilde{t}\,,\quad r=\sqrt{\tilde{r}^2+a^2\sin^2\tilde{\vartheta}}\,,\quad \cos\vartheta=\frac{\tilde{r}}{r}\cos\tilde{\vartheta}\,,\quad \varphi=\tilde{\varphi}\,,
	\end{eqnarray}
	where $a$ is a constant parameter. Then, the aforementioned arbitrary functions can be trivially re-defined from being dependent on $(r,\vartheta)$ to $(\tilde{r},\tilde{\vartheta})$.  Since the Boyer-Lindquist coordinates reduce to the spherical coordinates in the limit $a=0$, we omit the tilde in the following, as the presence or absence of the parameter $a$ is sufficient to distinguish which set of coordinates is used.
	
	The metric tensor, which corresponds to the tetrad \eqref{eq: general axial tetrad} contains all the possible cross terms $(dt\, dr, dt \, d\varphi, dt\, d\vartheta, dr \,d\varphi, dr \,d\vartheta, d\vartheta \, d\varphi )$. In particular it includes the tetrads for metrics which have only the $dt\, d\varphi$, component as off diagonal component, which are obtained by enforcing the relations 
	\begin{align}
	H_{00} H_{01}-H_{10} H_{11}-H_{20} H_{21}-H_{30} H_{31} &= 0\,,\\
	-H_{00} H_{02}-H_{10} H_{12}+H_{20} H_{22}+H_{30} H_{32} &= 0\,,\\
	-H_{01} H_{02}-H_{11} H_{12}+H_{21} H_{22}+H_{31} H_{32} &= 0\,,\\
	H_{01} H_{03}-H_{11} H_{13}-H_{21} H_{23}-H_{31} H_{33} &= 0\,,\\
	-H_{02} H_{03}-H_{12} H_{13}+H_{22} H_{23}+H_{32} H_{33} &= 0\,,
	\end{align}
	between the tetrad components, leaving $11$ of the $16$ free functions in the tetrad to be determined. 
	
	Inspired by the spherically symmetric tetrad which solves the antisymmetric $f(T,B,\phi,X)$ field equations, see~\cite{Ferraro:2011ks,Bahamonde:2019jkf,Bahamonde:2019zea,Tamanini:2012hg,Krssak:2015oua}, and the need to obtain one, and exactly one, off diagonal term, the $dt\, d\varphi$ term, in the metric to include the tetrads of the Kerr metric in Boyer-Lindquist coordinates, we introduce the reduced axially symmetric tetrad by setting
	\begin{equation}\label{eq:H1choice}
	H_{01}=H_{02}=H_{20}=H_{33}=H_{30}=H_{10}=H_{21}=H_{22}=H_{13}=0\,,\quad H_{31}=\frac{H_{11}H_{12}}{H_{32}}\,,
	\end{equation}
	which results in 
	\begin{equation}
	h^A{}_\mu=   \left(
	\begin{array}{cccc}
	H_{00} & 0 & 0 &H_{03} \\
	0 & H_{11}\cos \varphi  & H_{12}\cos \varphi  & -H_{23}\sin \varphi \\
	0 & H_{11}\sin \varphi  & H_{12}\sin \varphi  & H_{23}\cos \varphi \\
	0 & H_{11}H_{12}/H_{32} & -H_{32} & 0 \\
	\end{array}
	\right)\,.\label{tetrad}
	\end{equation}
	It is clear that this choice is valid only if $H_{32}\neq 0$. Although this choice may inadvertently leave out some solutions, quite likely these tetrads will play an important role in the search for solutions to the antisymmetric field equations of $f(T,B,\phi,X)$ gravity in axial symmetry, since for them only one of the six antisymmetric field equations is non-vanishing, as we will see in Sec.~\ref{sec:feq}. Moreover, they accommodate the axially symmetric solution found in~\cite{Jarv:2019ctf}, which however does not include Kerr or Taub-NUT geometry as special cases, see Sec.~\ref{sec:1}. 
	
	The reduced tetrads contain the well known spherically symmetric solutions~\cite{Bahamonde:2019jkf,Bahamonde:2019zea,Tamanini:2012hg,Krssak:2015oua}. In spherical coordinates, the latter is obtained by setting
	\begin{eqnarray}\label{spherical}
	H_{00}=\sqrt{\mathcal{A}(r)}\,, \quad  H_{11}=\sqrt{\mathcal{B}(r)}\sin\vartheta\,,\quad H_{12}=\sqrt{\mathcal{C}(r)}\cos\vartheta\,, \quad H_{23}=H_{32}=\sqrt{\mathcal{C}(r)}\sin\vartheta\,,\quad H_{03}=0\,,
	\end{eqnarray}
	where the functions $\mathcal{A}(r)$ and $\mathcal{B}(r)$ are to be determined by solving the symmetric field equations, and $\mathcal{C}(r)$ can be set to $\mathcal{C}(r)=r^2$, giving $g_{22}=r^2$ and $g_{33}=r^2\sin\vartheta$, without losing generality.
	
	The metric generated by the six independent free functions of the reduced axially symmetric tetrads becomes
	\begin{equation}\label{metric0}
	ds^2=H_{00}^2dt^2-H_{11}^2 \left(\frac{H_{12}^2}{H_{32}^2}+1\right)dr^2-(H_{12}^2+H_{32}^2)d\vartheta^2-(H_{23}^2-H_{03}^2)d\varphi^2+2 H_{00} H_{03}dt\, d\varphi\,.
	\end{equation}
	Metrics of this form are of particular interest since they contain the Pleba\'{n}ski–Demia\'{n}ski class of spacetimes, which we will discuss in Section~\ref{sec:axsymmet}. We will use them in this article to find new solutions of the antisymmetric field equations of $f(T,B,\phi,X)$ gravity which include teleparallel generalisations of the Taub-NUT metric and of the weakly rotating Kerr metric. Indeed, the absence of Birkhoff's theorem in modified teleparallel gravity allows the existence of non-trivial spherically symmetric vacuum solutions beyond the Schwarzschild geometry \cite{Bahamonde:2019jkf}. Accordingly, it is expected that the family of axially symmetric spacetimes is characterised by a much richer structure than the one present in TEGR.
	
	As a side result, it is interesting to mention that it is possible to obtain a different branch from the reduced axially symmetric tetrad \eqref{tetrad} which trivially solves the $f(T)$ antisymmetric equations by having a tetrad giving a vanishing torsion scalar ($T=0$) if one chooses
	\begin{eqnarray}\label{spherical2}
	H_{00}=\sqrt{\mathcal{A}(r)}\,, \quad  H_{11}=\sin\vartheta\,,\quad H_{12}=r \sqrt{1-\mathcal{A}(r) \sin^2\vartheta }\,, \quad H_{23}=r\sin\vartheta\,,\quad H_{32}=r \sqrt{\mathcal{A}(r)} \sin\vartheta \,,\quad H_{03}=0\,.
	\end{eqnarray}
	This tetrad generates a spherically symmetric metric but turns out to be a particular case of ~\eqref{spherical} since the metric components are constraint by the expression $g_{11}=-1/g_{00}$. It is worthwhile to stress that the tetrad reproduced by the above functions provides a spherically symmetric metric but the teleparallel connection does not respect spherical symmetry ($L_\zeta \Gamma\neq0$), in virtue of the existence of the component $T^{\varphi}{}_{\vartheta\varphi}$ and the inequality $T^{\varphi}{}_{r\varphi}\neq T^{\vartheta}{}_{r\vartheta}$, unless we assume the trivial case $g_{11}=1$. For the tetrad~\eqref{spherical2}, the dynamics of any $f(T)$ theory reduces to TEGR (or GR) plus an effective cosmological constant $\Lambda_{\rm eff}$ for any choice of $f$, and thus fixing $\mathcal{A}(r)=1-2M/r+\Lambda_{\rm eff}r^2$, the tetrad \eqref{spherical2} solves all of the $f(T)$ field equations.
	
	Moving on to $f(T,B)$ gravity, the tetrad~\eqref{spherical2} does not solve all the antisymmetric equations since the boundary term $B$ is given by
	\begin{eqnarray}
	B(r)=\frac{r^2 \mathcal{A}''(r)+4 r \mathcal{A}'(r)+2 \mathcal{A}(r)-2}{r^2}\,,
	\end{eqnarray}
	thus in general is non-vanishing. The antisymmetric field equations for $f(T,B)$ gravity can be solved by demanding that the boundary term vanishes too. This leads to  $\mathcal{A}(r)=1-2M/r+Q^2/r^2+\Lambda_{\rm eff} r^2$ that is the Reissner-Nordstr\"om  metric with a cosmological constant. 
	
	\subsubsection{Solely axially symmetric branch}\label{sec:alternative}
	The second branch of axially symmetric Weitzenb\"ock tetrads is obtained by choosing the trivial Lie group homomorphism leading to the Lie algebra homomorphism $\lambda(\partial_\varphi) = 0$, which implies that the tetrad components are simply independent of $\varphi$,
	\begin{align}\label{eq:2ndBranch}
	h^A{}_\mu = h^A{}_\mu(t,r,\vartheta)=\left(
	\begin{array}{cccc}
	H_{00} & H_{01} & H_{02} & H_{03} \\
	H_{10}  &H_{11}  & H_{12}   &H_{13}  \\
	H_{20}  & H_{21}  & H_{22} & H_{23}  \\
	H_{30} & H_{31} &H_{32} & H_{33} \\
	\end{array}
	\right)\,.
	\end{align}
	Generically in this branch, the tetrad has 16 components which only depend on $t,r$ and $\vartheta$. When we analyse this branch in Sec.~\ref{sec5.solely}, we will discuss time-dependent and time-independent tetrads.
	
	Since this tetrad is independent of $\varphi$, this branch does not include any of the spherically symmetric tetrads which solve the antisymmetric field equations of $f(T,B,\phi,X)$ gravity~\cite{Bahamonde:2019jkf,Bahamonde:2019zea,Tamanini:2012hg,Krssak:2015oua}. Hence it is a complete independent branch which leads to solutions of modified teleparallel theories of gravity which do not reduce to spherically symmetric teleparallel geometries in any case. Following the same idea as in the previous section, there is still some gauge freedom left. In order to eliminate all the cross terms in the metric except $dt \, d\varphi$, we choose the same combination of $H_{ij}$ as in~\eqref{eq:H1choice} but without setting $H_{20}=0$, to obtain the following tetrad
	\begin{equation}
	h^A{}_\mu=  \left(
	\begin{array}{cccc}
	H_{00} & 0 & 0 &H_{03} \\
	0 &H_{11} &H_{12} & 0 \\
	H_{20} & 0 & 0 &H_{23} \\
	0 & -\frac{H_{11}H_{12}}{H_{32}} & H_{32} & 0 \\
	\end{array}
	\right)\,,\label{tetradBrachB}
	\end{equation}
	and the metric
	\begin{eqnarray}
	ds^2&=&\left(H_{00}^2 -2 H_{20}^2\right)dt^2-  H_{11}^2\Big(1+\frac{H_{12}^2}{H_{32}^2}\Big)\, dr^2- (H_{12}^2+H_{32}^2)\, d\vartheta^2-\left(H_{23}^2-H_{03}^2\right)d\varphi^2\,\nonumber\\
	&&+2  (H_{00} H_{03}-H_{20} H_{23})\, dt \, d\varphi\,.
	\end{eqnarray}
	As for the first branch this choice is very convenient regarding the antisymmetric field equations since there one finds that there is only left which is $E_{[23]}$. 
	
	Due to the impossibility to connect these axially symmetric tetrads in a certain limit to spherically symmetric tetrads solving the antisymmetric field equations, we focus in what follows on finding solutions on the basis of the first branch.  Before we discuss some further details on the second branch and examples from the literature belonging to the second branch in section \ref{sec:2ndbranch}, we now discuss how the two branches are related and need to be understood in the context of the gauge freedom (i.e.\ local Lorentz transformations) in teleparallel gravity. 
	
	\subsubsection{Relationship between the branches}
	In teleparallel gravity, the fundamental variables are the components of a tetrad $h^A{}_\mu$ and a flat spin connection $\omega^A{}_{B\mu}$, or a tetrad $h^A{}_\mu$ and the spin connection generating local Lorentz transformation, $\Lambda^A{}_B$.
	
	From the symmetry arguments, in axial symmetry we find that the pairs $(h^A{}_\mu,\omega^A{}_{B\mu}) = (h_1^A{}_\mu,0)$, where $h_1^A{}_\mu$ is given by \eqref{eq: general axial tetrad}, and $(h^A{}_\mu,\omega^A{}_{B\mu}) = (h_2^A{}_\mu,0)$, where $h_2^A{}_\mu$ is given by \eqref{eq:2ndBranch}, define inequivalent axially symmetric teleparallel geometries.
	
	Although, both tetrads $h_1^A{}_\mu$ and $h_2^A{}_\mu$ are related by the local Lorentz transformation
	\begin{equation}
	\Lambda^A{}_B = \begin{pmatrix}
	1 & 0 & 0 & 0\\
	0 & \cos\varphi & \sin\varphi & 0\\
	0 & -\sin\varphi & \cos\varphi & 0\\
	0 & 0 & 0 & 1
	\end{pmatrix}\,,
	\end{equation}
	via the relation $h_2^A{}_\mu = \Lambda^A{}_B h_1^B{}_\mu$, in the covariant approach to teleparallel gravity this Lorentz transformation of the tetrads induces a spin connection, see \eqref{lortrans}, which has non-vanishing components
	\begin{equation}\label{eq:scaxsymm}
	\omega'^1{}_{2\varphi} = -\omega'^2{}_{1\varphi} = -1\,.
	\end{equation}
	Thus, the tuple $(h_1^A{}_\mu,0)$ is equivalent to the tuple $(h_2^A{}_\mu, \omega'^A{}_{B\mu})$ but not to the tuple $(h_2^A{}_\mu,0)$. In this sense the axially symmetric tetrads of the two branches here can either be considered as inequivalent Weitzenb\"ock tetrads, or as tetrads with non-vanishing spin connection.

	\section{Finding solutions to the antisymmetric field equations: the regular branch}\label{ssec:reular_branch}
	As previously mentioned, the relevance of the first branch of axially symmetric Weitzenb\"ock tetrads is highlighted in virtue of its compatibility with the spherically symmetric limit, which is why we study it here in detail. 
	
	We consider the well-known stationary axially symmetric Pleba\'{n}ski–Demia\'{n}ski class of metrics that are vacuum solutions of the Einstein equations and determine their reduced axially symmetric tetrad \eqref{tetrad}. As we will see, by doing so we fix five of their six free components. The remaining component must be determined from the antisymmetric field equations of the $f(T,B,\phi,X)$ theories. 
	
	We then search for solutions of the antisymmetric field equations of $f(T,B,\phi,X)$-gravity and find that using the reduced axially symmetric tetrad as ansatz, all but one antisymmetric field equation is solved. The remaining antisymmetric equation fixes one of the six tetrad components, while the remaining ones need to be determined by the symmetric field equations.
	
	We solve this last equation for generalisations of the tetrad of special subclasses of the Pleba\'{n}ski–Demia\'{n}ski metrics, such as the Taub-NUT and the Kerr metric for a slowly rotating black hole.
	
	\subsection{Pleba\'{n}ski–Demia\'{n}ski metric and its tetrads}\label{sec:axsymmet}
	The class of stationary axially symmetric algebraic type D vacuum solutions of GR (or TEGR) can be described by the Pleba\'{n}ski–Demia\'{n}ski metric with vanishing electromagnetic charges and cosmological constant. Their line element acquires the following form in Boyer-Lindquist type coordinates \cite{Plebanski:1976gy,Griffiths:2009dfa}:
	\begin{eqnarray}
	ds^{2}=\frac{1}{\Omega^{2}}\left\{\frac{\mathcal{Q}\;}{\varrho^2}\left[dt+\left(a\sin^2\vartheta+2b\left(\chi -\cos\vartheta\right)\right)d\varphi\right]^{2}-\frac{\varrho^2}{\mathcal{Q\;}}dr^2-\frac{\varrho^2}{\mathcal{P}\;}d\vartheta^2
	\right.
	\nonumber\\
	\left.
	-\,\frac{\mathcal{P}}{\varrho^2}\sin^2{\vartheta}\left[adt+\left(r^2+a^2+b^2+2\chi ab\right)d\varphi\right]^{2}\right\}\,,
	\label{PebDem}
	\end{eqnarray}
	with:
	\begin{eqnarray}
	\Omega&=&1-\frac{\alpha}{\gamma}\left(a\cos\vartheta+b\right)r\,,\\
	\varrho^2&=&\left(a\cos\vartheta+b\right)^2+r^2\,,\\
	\mathcal{Q}&=&k\,\gamma^2-2Mr+\epsilon\,r^2-2\alpha\frac{n}{\gamma}r^3-k\,\alpha^2r^4\,,\\
	\mathcal{P}&=&1+\left(4kab\alpha^{2}-2\alpha M\frac{a}{\gamma}\right)\cos\vartheta+k\,\alpha^{2}a^2\cos^2\vartheta\,,
	\end{eqnarray}
	where the constants $k,\epsilon$ and $n$ are defined as follows
	\begin{eqnarray}
	k&=&\frac{1+2b\alpha M/\gamma}{3b^2\alpha^2+\gamma^2/(a^2-b^2)}\,,\\
	\epsilon&=&\frac{k\gamma^2}{a^2-b^2}+4\alpha M\frac{b}{\gamma}-k\alpha^2\left(a^2+3b^2\right)\,,\\
	n&=&\frac{bk\gamma^2}{a^2-b^2}-\alpha M\frac{a^2-b^2}{\gamma}+bk\alpha^2\left(a^2-b^2\right)\,.
	\end{eqnarray}
	It includes four parameters $M$, $a$, $b$, and $\alpha$ representing the mass, angular momentum (per unit mass), NUT charge, and acceleration, respectively. In addition, the parameter $\chi$ sets the distribution of axial singularities\footnote{Note that despite the presence of a coordinate singularity on the polar axis, the axially symmetric Taub-NUT spacetime is geodesically complete and for the case $|\chi| \leq 1$ it does not lead to observable violations of causality in free-falling frames \cite{Clement:2015cxa}.}, whereas $\gamma$ represents a remaining scaling freedom for non-vanishing values of $a$ and $b$, which provides the twist of the underlying congruence of trajectories defined by the two different principal null directions of the solution \cite{Manko:2005nm,Griffiths:2005se}.
	Thereby, this family of solutions constitutes the natural generalisation of the special cases of Kerr, Taub-NUT, and C-metric spacetimes and describes the gravitational field generated by a uniformly accelerating and rotating black hole endowed with a NUT charge. 
	
	In terms of the metric~\eqref{metric0}, we can reproduce the Pleba\'{n}ski–Demia\'{n}ski spacetime~\eqref{PebDem} by setting the $H_{ij}$ functions in \eqref{tetrad} as follows:
	\begin{eqnarray}
	H_{00}&=&\frac{1}{\varrho\,\Omega}\sqrt{\mathcal{Q}-a^2\mathcal{P}\sin^2\vartheta}\,,\quad H_{11}=\sqrt{\frac{\mathcal{P}}{\mathcal{Q}}}\,H_{32}\,,\quad H_{12}=\sqrt{\frac{\varrho^2}{\mathcal{P}\,\Omega^2}-H_{32}^{2}}\,,\\
	H_{23}&=&\sqrt{\frac{\mathcal{P}\left(r^2+a^2+b^2+2\chi ab\right)^2\sin^2\vartheta-\mathcal{Q}\left(a\sin^{2}\vartheta+2b\left(\chi-\cos\vartheta\right)\right)^2}{\varrho^2\,\Omega^2}+H_{03}^{2}}\,,\\
	H_{03}&=&\frac{1}{\varrho\,\Omega\sqrt{\mathcal{Q}-a^2\mathcal{P}\sin^2\vartheta}}\left[\mathcal{Q}\left(a\sin^2\vartheta+2b\left(\chi-\cos\vartheta\right)\right)-a\mathcal{P}\sin^2\vartheta\left(r^2+a^2+b^2+2\chi ab\right)\right]\,.
	\end{eqnarray}
	
\begin{figure}[H]
		\centering
		\includegraphics[scale=1.2]{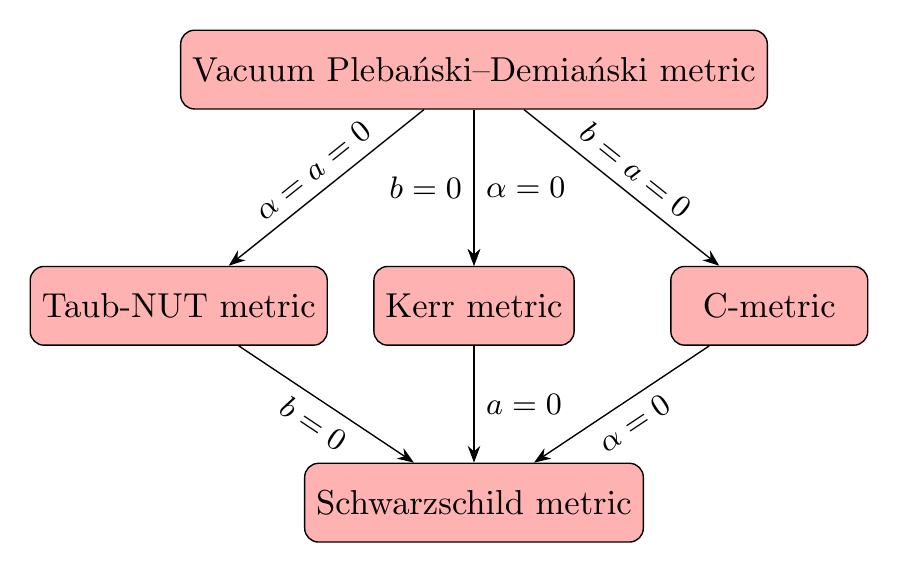}
		\caption{Relationship between the vacuum Pleba\'{n}ski–Demia\'{n}ski metric and other spacetimes.}
		\label{fig:PlebDem}
	\end{figure}

	Then, the well-known special cases included in the Pleba\'{n}ski–Demia\'{n}ski solution can be straightforwardly recovered in the following way, see also Figure \ref{fig:PlebDem}:
	\begin{enumerate}[i)]
		\item Kerr metric:
		\begin{eqnarray}
		H_{00}&=&\sqrt{1-\frac{2Mr}{\Sigma}}\,,\ \ H_{11}=\frac{H_{32}}{\sqrt{\Delta}}\,, \ \ H_{12}=\sqrt{\Sigma-H_{32}^2}\,,\label{kerrA}\\
		H_{23}&=&\sqrt{\sin^2\vartheta\left(\frac{2a^{2}Mr\sin^{2}\vartheta}{\Sigma}+a^2+r^2\right)+H_{03}^2}\,,\quad H_{03}=-\frac{2aMr\sin ^2\vartheta}{\sqrt{\Sigma(\Sigma-2Mr)}} \,,\label{kerrB}
		\end{eqnarray}
		where
		\begin{eqnarray}
		\label{eq:Sigma Delta def}
		\Omega&=&\mathcal{P}=1\,, \qquad
		\Sigma\equiv\varrho^{2}=r^2+a^2\cos^2\vartheta \,, \qquad \Delta\equiv\mathcal{Q}=\epsilon\,r^2-2Mr+k\gamma^{2}\,,\label{kerrC}\\
		b&=&\alpha=0\,, \qquad k\gamma^{2}=a^2\,, \qquad
		\epsilon=1\,, \qquad n=0\,.
		\end{eqnarray} 
		\item Taub-NUT metric:
		\begin{eqnarray}
		H_{00}&=&\sqrt{1-\frac{2\left(Mr+b^2\right)}{\varrho^2}}\,,\ \ H_{11}=\frac{H_{32}}{\sqrt{\mathcal{Q}}}\,, \ \ H_{12}=\sqrt{\varrho^2-H_{32}^2}\,,\\
		H_{23}&=&\varrho\sin\vartheta\,,\quad H_{03}=2b\left(\chi-\cos\vartheta\right)\sqrt{1-\frac{2\left(Mr+b^2\right)}{\varrho^2}}\,,\label{TN}
		\end{eqnarray}
		where
		\begin{eqnarray}
		\Omega&=&\mathcal{P}=1\,, \qquad
		\varrho^{2}=r^2+b^2\,, \qquad \mathcal{Q}=\epsilon\,r^2-2Mr+k\gamma^{2}\,,\\
		a&=&\alpha=0\,, \qquad k\gamma^{2}=-\,b^2\,, \qquad
		\epsilon=1\,, \qquad n=b\,.
		\end{eqnarray}
		\item C-metric:
		\begin{eqnarray}
		H_{00}&=&\frac{1}{r\Omega}\sqrt{\mathcal{Q}}\,,\quad\ \ H_{11}=H_{32} \sqrt{\frac{\mathcal{P}}{\mathcal{Q}}}\,,\quad H_{12}=\sqrt{\frac{r^2}{\Omega^2\mathcal{P}}- H_{32}^2}\,,\\
		H_{23}&=&\frac{r\sin\vartheta\sqrt{\mathcal{P}}}{\Omega}\,,\quad H_{03}=0\,,\label{CM}
		\end{eqnarray}
		where
		\begin{eqnarray}
		\Omega&=&1-\alpha r\cos\vartheta\,, \qquad \mathcal{P}=1-2\alpha M\cos\vartheta\,, \qquad
		\varrho^{2}=r^2\,, \qquad \mathcal{Q}=-2M r+\epsilon\,r^2-2\alpha\frac{n}{\gamma}r^3-k\,\alpha^2r^4\,,\label{CmetricFunc}\\
		a&=&b=0\,, \qquad k=1\,, \qquad
		\epsilon=1\,, \qquad n=-\,\alpha\gamma M\,,
		\end{eqnarray}
		and setting the remaining scaling factor $\gamma=a$.
		\item Schwarzschild metric:
		\begin{eqnarray}
		H_{00}&=&\sqrt{1-\frac{2M}{r}}\,,\ \ H_{11}=\frac{H_{32}}{\sqrt{\mathcal{Q}}}\,, \ \ H_{12}=\sqrt{r^2-H_{32}^2}\,,\label{SChwA}\\
		H_{23}&=&r\sin\vartheta\,,\quad H_{03}=0 \label{SChwB}\,,
		\end{eqnarray}
		where
		\begin{eqnarray}
		\Omega&=&\mathcal{P}=1\,, \qquad
		\varrho^{2}=r^2 \,, \qquad \mathcal{Q}=\epsilon\,r^2-2Mr+k\gamma^{2}\,,\\
		a&=&b=\alpha=0\,, \qquad k\gamma^{2}=0\,, \qquad
		\epsilon=1\,, \qquad n=0\,.
		\end{eqnarray}
	\end{enumerate}
	
	As can be seen, in all the cases we still have one extra function $H_{32}$ that does not influence the metric and must be obtained by solving the antisymmetric components of the field equations \eqref{fieldeq}. Then, the tetrad is what is called in the literature a good tetrad. 
	
	\subsection{The antisymmetric field equations}\label{sec:feq}
	If we substitute the reduced axially symmetric tetrad~\eqref{tetrad} into the $f(T,B,\phi,X)$ field equations \eqref{fieldeq}, we find that the only non-vanishing ones are the diagonal entries $E_{11},\, E_{22}, \, E_{33}, \, E_{44}$, and the off-diagonal entries $E_{14}, E_{32}, E_{23}$. Splitting the equations into symmetric and antisymetric part one realises that only one equation $E_{[23]}$ is non-zero. Thus, since the tetrad~\eqref{tetrad} contains one additional free function compared to the non-vanishing metric components (they fix 5 of 6 free functions in the tetrad), there is consistently one extra antisymmetric field equation to fix this tetrad component.
	
	If one is able to solve this equation, then, we would be able to find a good tetrad for the most general axially symmetric case. From~\eqref{antitotal}, $E_{[23]}$  then turns out to be
	\begin{equation}
	\frac{1}{2}\left[ (f_{T,\vartheta}+f_{B,\vartheta}) \underset{Q_\vartheta}{\underbrace{\left(\frac{H_{00,r}}{H_{00}}-\frac{H_{11}}{H_{23}}+\frac{H_{23,r}}{H_{23}}\right)}} 
	+  (f_{T,r}+f_{B,r}) \underset{Q_r}{\underbrace{\left(\frac{H_{12}-H_{23,\vartheta}}{H_{23}}-\frac{H_{00,\vartheta}}{H_{00}}\right)}}\right]=0\,,\label{antiGG}
	\end{equation}
	where $Q_\vartheta$, $Q_r$  were introduced as a shorthand notation and commas denote differentiation, i.e. 
	\begin{align}
	f_{T,\vartheta}&=\partial f_T/\partial \vartheta=f_{TT}T_{,\vartheta}+f_{TB}B_{,\vartheta}+f_{T\phi}\phi_{,\vartheta}+f_{TX}X_{,\vartheta}\,,\\
	f_{T,r}&=\partial f_T/\partial r=f_{TT}T_{,r}+f_{TB}B_{,r}+f_{T\phi}\phi_{,r}+f_{TX}X_{,r}\,,\\
	f_{B,\vartheta}&=\partial f_B/\partial \vartheta=f_{BT}T_{,\vartheta}+f_{BB}B_{,\vartheta}+f_{B\phi}\phi_{,\vartheta}+f_{BX}X_{,\vartheta}\,,\\
	f_{B,r}&=\partial f_B/\partial r=f_{BT}T_{,r}+f_{BB}B_{,r}+f_{B\phi}\phi_{,r}+f_{BX}X_{,r}\,.
	\end{align}
	Equation \eqref{antiGG} is difficult to solve in general for one of the functions $H_{ij}$ since all of them can depend on $r$ and $\vartheta$. One can easily notice that in the case where $f_T=-f_B$ which gives $f=f(-T+B,\phi,X)=f(\lc{R},\phi,X)$, the antisymmetric equations are satisfied  identically. This was expected since this case is an extension of the well-known $f(\lc{R})$ gravity. In addition, for $f(T,B)$ gravity, if one has the special case where both the scalar torsion and boundary term are constants ($T_{,r}=B_{,r}=T_{,\vartheta}=B_{,\vartheta}=0$), the above antisymmetric equation will be trivially satisfied and the dynamics of this theory will be just TEGR (or GR) plus a cosmological constant.
	
	By replacing the derivatives explicitly, one obtains that the remaining antisymmetric equation~\eqref{antiGG} becomes 
	\begin{eqnarray}
	0&=&  \bigg( (f_{TT}+f_{TB})T_{,\vartheta} + (f_{BT}+f_{BB})B_{,\vartheta} + (f_{T\phi}+f_{B\phi})\phi_{,\vartheta} + (f_{TX}+f_{BX})X_{,\vartheta} \bigg)Q_\vartheta\nonumber\\
	&& + \bigg( (f_{TT}+f_{TB})T_{,r} + (f_{BT}+f_{BB})B_{,r} + (f_{T\phi}+f_{B\phi})\phi_{,r} + (f_{TX}+f_{BX})X_{,r} \bigg)Q_r\,.\label{antiG}
	\end{eqnarray}
	
	This equation is still implicit since one would need to replace $T$ and $B$ (see Eq.~\eqref{Tgeneral}-\eqref{Bgeneral}) to find it in its complete form. One can then have different options to solve this equation. Some important cases are:
	\begin{enumerate}
		\item $Q_\vartheta = 0$ and $Q_r = 0$: This case leads to universal solutions for any choice of $f$ and any form of $T,B$ and $\phi$ (see~Sec.~\ref{sec:1}). 
		\item $f_{T,\vartheta}+f_{B,\vartheta} = 0$ and $Q_r = 0$ (see~Sec.~\ref{sec:2}). 
		\item   $f_{T,r}+f_{B,r} = 0$ and $Q_\vartheta = 0$ (see~Sec.~\ref{sec:strange}). 
		\item $(f_{T,r}+f_{B,r})Q_r \neq0$, $(f_{T,\vartheta}+f_{B,\vartheta})Q_{\vartheta}\neq0$: This case is the most general one. To solve the antisymmetric equation one would need to replace the value of the scalars and then solve the equation for any of the functions $H_{ij}$. This case would give the most general good axially symmetric tetrad, but the equation is very involved and difficult to treat. The Kerr metric is part of this case (see~Sec.~\ref{sec:case6}).
	\end{enumerate}

	\subsection{Solving the antisymmetric equation}\label{ref:examples}
	In the following, we will study the antisymmetric field equations for the different cases directly to find the solutions. Indeed, considering the highly nonlinear character of these equations, the complexity for axially symmetric configurations turns out to be much more involved compared to their spherical symmetric counterparts and requires a deeper examination. Accordingly, we now discuss the four cases framed in teleparallel theories beyond TEGR which are inequivalent to $f(\lc{R},\phi,X)$ theories.
	
	\subsubsection{Case 1: $Q_{\vartheta}=Q_{r}=0$}\label{sec:1}
	In this case, it is easy to solve the antisymmetric equations since $Q_{\vartheta}=Q_{r}=0$ (see~\eqref{anti23}). We find the following solutions
	\begin{eqnarray}
	H_{11}= \frac{H_{00,r}H_{23}}{H_{00}}+H_{23,r}\,,\quad H_{12}= \frac{H_{00,\vartheta}H_{23}}{H_{00}}+H_{23,\vartheta}\,.
	\end{eqnarray}
	By choosing these two functions, one finds a tetrad which explicitly reads as
	\begin{eqnarray}\label{goodtetrad1}
	h^A{}_\mu&=&\left(
	\begin{array}{cccc}
	H_{00} & 0 & 0 &H_{03} \\
	0 &  \left(\frac{H_{00,r}H_{23}}{H_{00}}+H_{23,r}\right)\cos \varphi  &  \left(\frac{H_{00,\vartheta}H_{23}}{H_{00}}+H_{23,\vartheta}\right)\cos \varphi  & -H_{23}\sin \varphi \\
	0 &  \left(\frac{H_{00,r}H_{23}}{H_{00}}+H_{23,r}\right)\sin \varphi  &  \left(\frac{H_{00,\vartheta}H_{23}}{H_{00}}+H_{23,\vartheta}\right)\sin \varphi  &H_{23} \cos \varphi  \\
	0 &H_{32}^{-1}\left(\frac{H_{00,\vartheta}H_{23}}{H_{00}}+H_{23,\vartheta}\right) \left(\frac{H_{00,r}H_{23}}{H_{00}}+H_{23,r}\right) & -H_{32} & 0 \\
	\end{array}
	\right) \,.
	\end{eqnarray}
	This tetrad solves the antisymmetric field equation universally for all $f(T,B,\phi,X)$.  In the above tetrad, all the functions can depend on both $r$ and $\vartheta$ and while the form of the torsion scalar $T$ and the boundary term $B$ are very cumbersome (see~\eqref{Tgeneral}-\eqref{Bgeneral}). This case constrains the metric and does not contain either Kerr, Taub-NUT, or the C-metric.
	However, this case generalises the good tetrad found in~\cite{Jarv:2019ctf}. This can be seen by performing a Lorentz transformation $h'{}^A{}_\mu=\Lambda^A{}_B h{}^B{}_\mu$ with the Lorentz matrix being 
	\begin{equation}\label{Lorentz1}
	\Lambda^A{}_B=\left(
	\begin{array}{cccc}
	1 & 0 & 0 & 0 \\
	0 & 0 & 0 & 1 \\
	0 & 1 & 0 & 0 \\
	0 & 0 & 1 & 0 \\
	\end{array}
	\right)\,.
	\end{equation}
	
	Note that the Lorentz transformation~\eqref{Lorentz1} just cyclically relabels the local axes and does not introduce any spin connection ($\omega'{}^{A}{}_{B\mu}=0$), since it is constant. The special case of the good tetrad found in~\cite{Jarv:2019ctf}  (see Eq.~(32) there), can be recovered by doing this local Lorentz transformation and then setting
	\begin{eqnarray}\label{eqLaur}
	H_{00}&=&\mathcal{A}(r,\vartheta) \,, \quad H_{03}=-\mathcal{A}(r,\vartheta)\mathcal{W}(r,\vartheta) \,, \quad H_{11}=-\frac{(r-M) \sin\vartheta}{\mathcal{A}(r,\vartheta)\sqrt{\Delta}} \,, \quad H_{12}=-\frac{\sqrt{\Delta} \cos\vartheta}{\mathcal{A}(r,\vartheta)} \,, \\
	H_{23}&=&-\frac{\sqrt{\Delta}\sin\vartheta}{\mathcal{A}(r,\vartheta)} \,, \quad H_{32}=-\frac{(r-M)\sin\vartheta}{\mathcal{A}(r,\vartheta)}\,,\label{eqLaur2}
	\end{eqnarray}
	where $\mathcal{A}(r,\vartheta)$ and $\mathcal{W}(r,\vartheta)$ are some functions of $r,\vartheta$ (to be determined by the symmetric equations), and $\Delta$ was defined in Eq.~\eqref{eq:Sigma Delta def}.
	
	One interesting remark about the tetrad given by~\eqref{eqLaur}-\eqref{eqLaur2} is that if one sets the two arbitrary functions to be
	\begin{eqnarray}
	\mathcal{A}(r,\vartheta)=\frac{\sqrt[4]{M^2-a^2}}{\sqrt{K_1 \arctan \left(\frac{r-M}{\sqrt{a^2-M^2}}\right)-2 K_2 \sqrt{M^2-a^2}}}\,,\quad \mathcal{W}(r,\vartheta)=K_3+K_1\cos\vartheta\,,
	\end{eqnarray}
	where $K_i$ are constants, one notices that $\lc{R}=0$, thus, the corresponding metric is an exact solution of GR (or TEGR). Then, for the above form of the functions, the boundary term $B$ and the scalar torsion $T$ are equal. Moreover, the scalar torsion and the boundary term are different to zero, unless $K_1=0$. 
	
	\subsubsection{Case 2: $Q_r=0$ and $f_{T,\vartheta}+f_{B,\vartheta} = 0$ - Taub-NUT-like metric}\label{sec:2}
	In this section, we are going to study the special case where we have a metric with a Taub-NUT-like form. To do so we choose
	\begin{eqnarray}
	H_{23}&=&\sin \vartheta \sqrt{\mathcal{C}(r)}\,,\quad  H_{03}=\sqrt{\mathcal{A}(r)}\mathcal{D}(r,\vartheta)\,,\quad H_{00}=\sqrt{\mathcal{A}(r)}\,,\\
	H_{11}&=&H_{32}(r,\vartheta)\sqrt{\frac{\mathcal{B}(r )}{\mathcal{C}(r)}}\,,\quad H_{12}=\sqrt{\mathcal{C}(r)-H_{32}(r,\vartheta)^2}\,.
	\end{eqnarray}
	The Taub-NUT metric, which is a specific case \eqref{TN}, of the Pleba\'{n}ski–Demia\'{n}ski metrics~\eqref{PebDem}, is recovered by choosing
	\begin{equation}
	\mathcal{A}(r)=1/\mathcal{B}(r)=\frac{(r-r_{+})(r-r_{-})}{r^2+b^2}\,, \quad \mathcal{C}(r)=r^2+b^2\,,\quad \mathcal{D}(r,\vartheta)=2b \cos\vartheta\,,\quad r_{\pm}=M\pm\sqrt{M^2+b^2}\,,
	\end{equation}
	where $b$ is the Taub-NUT parameter and $M$ the mass. The Taub-NUT parameter plays the role of a gravitomagnetic monopole moment, in virtue of its physical effect on the trajectories of test particles minimally coupled to the connection \cite{Zimmerman:1989kv,LyndenBell:1996xj}. It is not asymptotically flat, at large distances it provides a non-vanishing component $g_{03} = 2b\left(\chi-\cos\vartheta\right)$ which acts as a gravitomagnetic potential \cite{dowker1974nut}.
	
	The antisymmetric equation~\eqref{antiG}  for this case becomes
	\begin{eqnarray}
	0&=&\frac{1}{4} (f_{T,\vartheta}+f_{B,\vartheta}) \left(\frac{\mathcal{A}'(r)}{\mathcal{A}(r)}+\frac{\mathcal{C}'(r)}{\mathcal{C}(r)}-2\sqrt{\frac{\mathcal{B}(r)   }{\mathcal{C}(r)}}H_{32}(r,\vartheta)\csc \vartheta\right)\nonumber\\ 
	&&+\frac{1 }{2}(f_{T,r}+f_{B,r})\left(\csc\vartheta  \sqrt{1-\frac{H_{32}(r,\vartheta )^2}{\mathcal{C}(r)}}-\cot\vartheta \right)\,.\label{anti23}
	\end{eqnarray}
	Even though this equation is still implicit, the second parenthesis vanishes for
	\begin{equation}
	H_{32}(r,\vartheta)=\sqrt{\mathcal{C}(r)}\sin\vartheta=H_{23}(r,\vartheta)\,.
	\end{equation}
	With this choice, \eqref{anti23} is not fully satisfied since we still have the term multiplied by $f_{T,\vartheta}+f_{B,\vartheta}$ remaining. The torsion scalar and the boundary term, see~\eqref{Tgeneral}-\eqref{Bgeneral}, for this case are
	\begin{eqnarray}
	T(r,\vartheta)&=&\mathcal{A}(r)\csc ^2\vartheta \left(\frac{ \mathcal{D}_{,r}^2}{2 \mathcal{B}(r) \mathcal{C}(r)}+\frac{ \mathcal{D}_{,\vartheta}^2}{2 \mathcal{C}(r)^2}\right)+\mathcal{C}'(r) \left(\frac{\mathcal{A}'(r)}{\mathcal{A}(r) \mathcal{B}(r) \mathcal{C}(r)}-\frac{1}{\sqrt{\mathcal{B}(r)} \mathcal{C}(r)^{3/2}}\right)\nonumber\\
	&&-\frac{\mathcal{A}'(r)}{\mathcal{A}(r) \sqrt{\mathcal{B}(r)\mathcal{C}(r)}}+\frac{\mathcal{C}(r)'{}^2}{2 \mathcal{B}(r) \mathcal{C}(r)^2}+\frac{2}{\mathcal{C}(r)}\,,\\
	B(r)&=&-\frac{1}{2\mathcal{A}(r)^2\mathcal{B}(r)^2}\Big[\mathcal{B}(r) \mathcal{A}'(r)^2+\mathcal{A}(r) \left(\mathcal{A}'(r) \mathcal{B}'(r)+\frac{4 \mathcal{B}(r)^{3/2} \mathcal{A}'(r)}{\sqrt{\mathcal{C}(r)}}+\mathcal{B}(r) \left(-2 \mathcal{A}''(r)-\frac{4 \mathcal{A}'(r) \mathcal{C}'(r)}{\mathcal{C}(r)}\right)\right)\nonumber\\
	&&+\mathcal{A}(r)^2 \left(\frac{2 \mathcal{B}'(r) \mathcal{C}'(r)}{\mathcal{C}(r)}-\frac{4 \mathcal{B}(r) \mathcal{C}''(r)}{\mathcal{C}(r)}+\frac{4 \mathcal{B}(r)^{3/2} \mathcal{C}'(r)}{\mathcal{C}(r)^{3/2}}\right)\Big]\,.
	\end{eqnarray}
	One first notices that the boundary term only depends on the radial coordinate. If one then considers theories of the type $f(T,B,\phi,X)=T+\tilde{f}_1(B,\phi,X)+\tilde{f}_2(\lc{R},\phi,X)$ with $\phi=\phi(r)$, the antisymmetric field equations are immediately satisfied. Second, keeping $f$ general we can set
	\begin{equation}
	\mathcal{  D}(r,\vartheta)=C_1 \cos \vartheta +C_2\,,
	\end{equation}
	which implies that also the torsion scalar only depends on $r$. Thus, now $T=T(r)$ and $B=B(r)$ and the remaining part of the antisymmetric equation~\eqref{anti23} becomes
	\begin{eqnarray}
	0&=&\left[\mathcal{A}(r) \left(2 \sqrt{\mathcal{B}(r)\mathcal{C}(r)}-\mathcal{C}'(r)\right)-\mathcal{C}(r) \mathcal{A}'(r)\right] \times \Big[\mathcal{B}(r) \phi_{,\vartheta} \Big(\mathcal{C}(r) (f_{B\phi}+f_{T\phi})+ \epsilon\, \phi_{,\vartheta\vartheta} (f_{BX}+f_{TX})\Big)\nonumber\\
	&&+ \epsilon\,\mathcal{C}(r) \phi_{,r} \phi_{,r\vartheta} (f_{BX}+f_{TX})\Big]\,.
	\end{eqnarray}
	This equation has different subclasses of solutions, which can be summarised as follows:
	\begin{enumerate}
		\item $\phi=\phi(r)$: this subclass assumes that the scalar field only depends on the radial coordinate.
		\item $f_{B\phi}+f_{T\phi}=f_{BX}+f_{TX}=0$ implying $f(T,B,\phi,X)=f_{1}(T,B)+f_2(-T+B,\phi,X)=f_{1}(T,B)+f_2(\lc{R},\phi,X)$: this subclass is only $f(T,B)$ plus a scalar-tensor theory based on the Ricci scalar.
		\item $\mathcal{B}(r)= \tfrac{\left(\mathcal{C}(r) \mathcal{A}'(r)+\mathcal{A}(r) \mathcal{C}'(r)\right)^2}{4 \mathcal{A}(r)^2 \mathcal{C}(r))}$ ($Q_\vartheta=0$): this subclass constraint the metric to a special one that does not have the Taub-NUT metric as a special case. This case is identical to the case where $Q_r=Q_\vartheta=0$ with some extra conditions in the metric.
	\end{enumerate}
	Let us now focus on the solutions within subclasses 1 and 2. For these cases, the following tetrad
	\begin{equation}
	h^A{}_\mu=\left(
	\begin{array}{cccc}
	\sqrt{\mathcal{A}(r)} & 0 & 0 & \sqrt{\mathcal{A}(r)} \left(C_2+C_1\cos \vartheta\right) \\
	0 & \sqrt{\mathcal{B}(r)} \sin \vartheta \cos \varphi & \sqrt{\mathcal{C}(r)}\cos \vartheta  \cos \varphi & - \sqrt{\mathcal{C}(r)}\sin \vartheta\sin \varphi  \\
	0 & \sqrt{\mathcal{B}(r)} \sin \vartheta \sin \varphi & \sqrt{\mathcal{C}(r)}\cos \vartheta \sin\varphi & \sqrt{\mathcal{C}(r)}\sin \vartheta \cos \varphi  \\
	0 & \sqrt{\mathcal{B}(r)} \cos \vartheta & -\sqrt{\mathcal{C}(r)}\sin \vartheta  & 0 \\
	\end{array}
	\right)\label{tetradgood}
	\end{equation}
	is a good tetrad that solves the antisymmetric field equations. This tetrad behaves very similarly to the spherically symmetric good tetrad (see Eq.~\eqref{spherical}), however, contains an extra term related to axial symmetry. As a special case, it is a tetrad of the Taub-NUT metric. The general metric associated with the above tetrad is
	\begin{eqnarray}\label{metrictaublike}
	ds^2&=&\mathcal{A}(r)dt^2 -\mathcal{B}(r)dr^2- \mathcal{C}(r)d\vartheta^2- \left[\mathcal{C}(r)\sin ^2\vartheta -\mathcal{A}(r) \left(C_1 \cos \vartheta +C_2\right)^2\right]d\varphi^2\nonumber\\
	&&+2  \mathcal{A}(r) \left(C_1 \cos \vartheta +C_2\right)dt\, d\varphi\,.
	\end{eqnarray}
	The torsion tensor given by the good axially symmetric tetrad~\eqref{tetradgood} can be decomposed as the standard spherically symmetric torsion piece plus an additional piece coming from axial symmetry, namely
	\begin{eqnarray}
	T_{\alpha\mu\nu}=T_{\alpha\mu\nu}^{(\rm spher)}+T_{\alpha\mu\nu}^{(\rm axi)}\,,
	\end{eqnarray}
	where the non-zero components of the axial part are
	\begin{eqnarray}
	T_{tr\varphi}^{(\rm axi)}&=&-T_{\varphi t r}^{(\rm axi)}=(C_2+C_1 \cos \vartheta )^{-1}T_{\varphi r \varphi}^{(\rm axi)}=\frac{1}{2} \mathcal{A}'(r) (C_2+C_1 \cos \vartheta )\,,\\
	T_{t\vartheta\varphi}^{(\rm axi)}&=&(C_2+C_1 \cos \vartheta )^{-1}T_{\varphi \vartheta \varphi}^{(\rm axi)}=-C_1\mathcal{A}(r)\sin\vartheta\,. 
	\end{eqnarray}
	
	We can also perform a Lorentz transformation for both the tetrad and the spin connection, with $\Lambda^A{}_B $ being,
	\begin{equation}\label{eq:spherlt1}
	\Lambda^A{}_B = \begin{pmatrix}
	1 & 0 & 0 & 0\\
	0 & \sin\vartheta\cos\varphi & \sin\vartheta\sin\varphi & \cos\vartheta\\
	0 & \cos\vartheta\cos\varphi & \cos\vartheta\sin\varphi & -\sin\vartheta\\
	0 & -\sin\varphi & \cos\varphi & 0
	\end{pmatrix}\,,
	\end{equation}
	which gives us that the transformed tetrad is
	\begin{equation}\label{tetradgood2}
	h'^A{}_\mu=\left(
	\begin{array}{cccc}
	\sqrt{\mathcal{A}(r)} & 0 & 0 & \sqrt{\mathcal{A}(r)} \left(C_2+C_1 \cos \vartheta\right) \\
	0 & \sqrt{\mathcal{B}(r)}  & 0 & 0\\
	0 & 0& \sqrt{\mathcal{C}(r)}& 0 \\
	0 & 0& 0  & \sqrt{\mathcal{C}(r)}\sin \vartheta \\
	\end{array}
	\right)\,,
	\end{equation}
	and then with a non-zero spin connection yielding
	\begin{equation}\label{eq:spiconspher1}
	\omega'^{r}{}_{\vartheta\vartheta}=-\omega'^{\vartheta}{}_{r\vartheta}=-1\,,\quad \omega'^{r}{}_{\varphi\varphi}=-\omega'^{\varphi}{}_{r\varphi}=-\sin\vartheta\,,\quad \omega'^{\vartheta}{}_{\varphi\varphi}=-\omega'^{\varphi}{}_{\vartheta\varphi}=-\cos\vartheta\,.
	\end{equation}
	It is then equivalent to use the tetrad~\eqref{tetradgood} with a vanishing spin connection than to use the tetrad-spin connection pair given by~\eqref{tetradgood2} and \eqref{eq:spiconspher1}. It is worth noticing that the above spin connection is exactly the same as the one reported in spherical symmetry in other papers~\cite{Hohmann:2019nat}.

	\subsubsection{Case 3: $Q_\vartheta=0$ and  $f_{T,r}+f_{B,r} = 0$}\label{sec:strange}
	There is another case where it is possible to find a good tetrad, namely, when $Q_\vartheta=0, f_{T,r}+f_{B,r}=0 $. In order to find a solution for this case, let us assume that the functions $H_{ij}$ are
	\begin{eqnarray}\label{Hijcases}
	H_{12}(r,\vartheta)=0\,,\quad H_{00}(r,\vartheta)=\mathcal{A}(\vartheta)\,,\quad 
	H_{03}(r,\vartheta)=\mathcal{B}(\vartheta)\,,\quad H_{32}(r,\vartheta)=\mathcal{C}(\vartheta)\,,\quad H_{23}(r,\vartheta)=\mathcal{D}_1(\vartheta)\mathcal{D}_2(r)\,,
	\end{eqnarray}
	and then since $Q_\vartheta=0$ we have $H_{11}(r,\vartheta)=\mathcal{D}_1(\vartheta )\mathcal{D}_{2}'(r)$. After assuming this, the torsion scalar becomes
	\begin{eqnarray}
	T(r,\vartheta)&=&\frac{-2 \mathcal{A}(\vartheta) \mathcal{B}(\vartheta) \mathcal{A}'(\vartheta) \mathcal{B}'(\vartheta)+\mathcal{B}(\vartheta)^2 \mathcal{A}'(\vartheta)^2+\mathcal{A}(\vartheta)^2 \mathcal{B}'(\vartheta)^2}{2 \mathcal{A}(\vartheta)^2 \mathcal{D}_1(\vartheta)^2 \mathcal{D}_2(r)^2 \mathcal{C}(\vartheta)^2}+\frac{8 \mathcal{A}(\vartheta) \mathcal{D}_1(\vartheta) \mathcal{A}'(\vartheta) \mathcal{D}_1'(\vartheta)+4 \mathcal{A}(\vartheta)^2 \mathcal{D}_1'(\vartheta)^2}{2 \mathcal{A}(\vartheta)^2 \mathcal{D}_1(\vartheta)^2 \mathcal{C}(\vartheta)^2}\,,\nonumber\\
	\end{eqnarray}
	and then if one further assumes
	\begin{equation}
	\mathcal{A}(\vartheta)=\mathcal{B}(\vartheta)\label{H11b}
	\end{equation}
	the scalar torsion will depend only on $\vartheta$ (i.e. $T_{,r}=0$). Moreover, the same happens for the boundary term ($B_{,r}=0$). Explicitly these quantities read as follows
	\begin{eqnarray}
	T(\vartheta)&=&\frac{2 \mathcal{D}_1'(\vartheta ) \left(2 \mathcal{D}_1(\vartheta ) \mathcal{B}'(\vartheta )+\mathcal{B}(\vartheta ) \mathcal{D}_1(\vartheta )\right)}{\mathcal{B}(\vartheta ) \mathcal{D}_1(\vartheta )^2\mathcal{C}(\vartheta )^2}\,,\\
	B(\vartheta)&=&\frac{1}{\mathcal{B}(\vartheta) \mathcal{D}_1(\vartheta)^2 \mathcal{C}(\vartheta)^3}\Big[4 \mathcal{D}_1(\vartheta) (\mathcal{C}(\vartheta) (2 \mathcal{B}'(\vartheta) \mathcal{D}_1'(\vartheta)+\mathcal{B}(\vartheta) \mathcal{D}_1''(\vartheta))\nonumber\\
	&&-\mathcal{B}(\vartheta) \mathcal{D}_1'(\vartheta) \mathcal{C}'(\vartheta))+\mathcal{D}_1(\vartheta)^2 \left(2 \mathcal{C}(\vartheta) \mathcal{B}''(\vartheta)-2 \mathcal{B}'(\vartheta) \mathcal{C}'(\vartheta)\right)+4 \mathcal{B}(\vartheta) \mathcal{C}(\vartheta) \mathcal{D}_1'(\vartheta)^2\Big]\,.\nonumber \\
	\end{eqnarray}
	After all these assumptions, the antisymmetric equation~\eqref{antiG} becomes
	\begin{eqnarray}
	0&=&Q_r \Big[\epsilon \,(f_{BX}+f_{TX}) \left(\mathcal{D}_1(\vartheta)^2 \mathcal{D}_2'(r))^3 \phi_{,\vartheta} \phi_{,r\vartheta}+\mathcal{C}(\vartheta)^2 \phi_{,r} \left(\mathcal{D}_2'(r)) \phi_{,rr}-\mathcal{D}_2''(r) \phi_{,r}\right)\right)\nonumber\\
	&&+\mathcal{D}_1(\vartheta)^2 \mathcal{C}(\vartheta)^2 \mathcal{D}_2'(r))^3 f_{B\phi} \phi_{,r}+\mathcal{D}_1(\vartheta)^2 \mathcal{C}(\vartheta)^2 \mathcal{D}_2'(r))^3 f_{T\phi} \phi_{,r}\Big]\,.
	\end{eqnarray}
	Similarly as the case studied before, we can have three different solutions. For $Q_{r}\neq 0$ and $f\neq f_1(T,B)+f_2(\lc{R},\phi,X)$, we require $\phi=\phi(\vartheta)$ to solve the above equation. Thus, when $\phi=\phi(\vartheta)$, the following tetrad
	\begin{equation}\label{goodtetrad3}
	h^A{}_\mu=\left(
	\begin{array}{cccc}
	\mathcal{B}(\vartheta) & 0 & 0 &\mathcal{B}(\vartheta) \\
	0 &\mathcal{D}_1(\vartheta)\mathcal{D}_{2}'(r) \cos \varphi  & 0 & -\mathcal{D}_1(\vartheta) \mathcal{D}_2(r) \sin \varphi  \\
	0 &\mathcal{D}_1(\vartheta)\mathcal{D}_{2}'(r) \sin \varphi   & 0 &\mathcal{D}_1(\vartheta)  \mathcal{D}_2(r) \cos \varphi  \\
	0 & 0 & -\mathcal{C}(\vartheta) & 0 \\
	\end{array}
	\right)
	\end{equation}
	solves the antisymmetric field equations, and then it is a good tetrad for $f(T,B,\phi,X)$ gravity. The metric corresponding to the above tetrad is
	\begin{equation}
	ds^2=\mathcal{B}(\vartheta)^2\, dt^2    - \mathcal{D}_1(\vartheta)^2 \mathcal{D}_{2}'(r)^2\, dr^2- \mathcal{C}(\vartheta)^2\, d\vartheta^2+ \left(\mathcal{B}(\vartheta)^2-\mathcal{D}_1(\vartheta)^2 \mathcal{D}_2(r)^2\right)d\varphi^2+2  \mathcal{B}(\vartheta)^2\, dt\, d\varphi\,.
	\end{equation} 
	If $\mathcal{B}(\vartheta)\neq \textrm{const.}$, this metric is never spherically symmetric since $g_{00}=g_{03}=\mathcal{B}(\vartheta)^2$.
	
	\subsubsection{Case 4: $(f_{T,r}+f_{B,r})Q_r \neq0$ and $(f_{T,\vartheta}+f_{B,\vartheta})Q_{\vartheta}\neq0$ -- Kerr metric and its perturbation} \label{sec:case6}
	This case is very involved since one must replace the form of the scalar torsion $T$ and $B$ (see~\eqref{Tgeneral} and \eqref{Bgeneral}) in the antisymmetric equation~\eqref{antiGG} and then solve the partial differential equation for one of the functions $H_{ij}$. If one is able to solve this antisymmetric equation directly, one could have fixed one function of the $H_{ij}$, and then, one would be able to obtain a good axially symmetric tetrad having five arbitrary functions $H_{ij}$. Then, one could use this tetrad to find solutions to the remaining field equations (symmetric part). One notices that the Kerr and the C-metric are part of this case. 
	
	Since the general case is very involved, we can first try to analyse the specific case and use the tetrad \eqref{kerrA}-\eqref{kerrB} for the Kerr metric. If we do this, we will only have one free function (for example $H_{32}$) that needs to be determined from the antisymmetric equation. Even though this is just a particular case and it would not give us the result needed to find a general good axially symmetric tetrad, it will be useful as a first step. Moreover, it can be noticed from the previous sections that the form of the good tetrad associated with Schwarzschild and the Taub-NUT metrics have a trivial generalisation to a general spherically symmetric and a Taub-NUT-like metric cases, respectively. Thus, if we are able to find a good tetrad for the Kerr case, this could give us a hint to tackle the general axially symmetric case.
	
	Another motivation to find this tetrad is related to the search of teleparallel perturbations of Kerr geometry, similar to the $f(T,B)$ perturbations of Schwarzschild geometry studied in \cite{Ruggiero:2015oka,Bahamonde:2019jkf,Bahamonde:2019zea,Bahamonde:2020bbc}. In perturbation theory around a TEGR solution, the antisymmetric field equations only contain the TEGR background tetrad and thus, a good tetrad for Kerr geometry could be used as a starting point for a perturbative analysis beyond TEGR as it has been
	done in curvature-based extensions of GR~\cite{Konno:2009kg, Yunes:2009hc, Pani:2009wy, Pani:2011gy, Ayzenberg:2014aka,Maselli:2015tta,Cano:2019ore}.
	
	Plugging the tetrad for the Kerr metric~\eqref{kerrA}-\eqref{kerrB} in the remaining antisymmetric field equation~\eqref{antiG} and trying to solve it for the undetermined tetrad component $H_{32}(r,\vartheta)$ turns out to be a difficult task. The main obstacle is, that in this case, all terms $(f_{T,r}+f_{B,r})Q_r \neq0$, $(f_{T,\vartheta}+f_{B,\vartheta})Q_{\vartheta}\neq0$ are non-vanishing, i.e. one has to deal with derivatives of the torsion scalar and the boundary term (see~\eqref{Tgeneral} and \eqref{Bgeneral}). Below we present a solution of the antisymmetric field equations for an expansion of the Kerr spacetime for small rotation parameter $a$, i.e. for slowly rotating black holes. This is the first step in the ongoing research project to solve the antisymmetric field equation analytically for Kerr geometry in $f(T,B,\phi,X)$ gravity. 
	
	First one notices that if
	\begin{equation}\label{H43is}
	H_{32}(r,\vartheta)=r \sin\vartheta\,,
	\end{equation}
	the antisymmetric equation~\eqref{antiG} is satisfied for either $a=0$ or $M=0$, but not when both are different from zero. For the $a=0$ case, one also needs to assume that $\phi=\phi(r)$ as an extra condition which is consistent with the fact this case is just Schwarzschild which is spherically symmetric. Moreover, for this case, when $a=0$ one recovers the standard spherically symmetric good tetrad with Schwarzschild metric components (which has $T\neq0$), see for example \cite{Bahamonde:2019zea}. Then, one can conclude that the general case when $a\neq0$ and $M\neq0$ must contain terms like $a^n M^p$ in the above function. Let us now assume that $a\ll 1$ and make an expansion around $a=0$. For this case, we can consider the following form of the function
	\begin{equation}
	H_{32}(r,\vartheta)=   r \sin\vartheta+a^2\, \mathcal{A}(r,\vartheta)+a^3\, \mathcal{B}(r,\vartheta)+\mathcal{O}(a^4)\,,\label{H43exp}
	\end{equation}
	and expand $a$ up to third order. For simplicity, we will only consider $f(T,B,\phi,X)=\tilde{f}(T,B)+f_2(\lc{R},\phi,X)$ since the case with the scalar field is even more involved. 
	
	Inserting the above function in~\eqref{antiG} and expanding up to second order we find the following differential equation for $\mathcal{A}$
	\begin{eqnarray}
	0&=&a^2(\tilde{f}_{TT}+2\tilde{f}_{TB}+\tilde{f}_{BB})\Big[2 \sin (2 \vartheta ) \left(\sqrt{r}-\mu\right) \Big\{4 r^{3/2} (\cos (2 \vartheta )-3) \left(\sqrt{r}-\mu\right) \mu^2 \mathcal{A}_{,\vartheta}\nonumber\\
	&&-4 \sin (2 \vartheta ) \Big(r^{3/2} \left(\sqrt{r}-\mu\right) \mu^2 \mathcal{A}_{,\vartheta\vartheta}+2 \sin \vartheta  \cos ^2\vartheta \left(5 r^{3/2} \mu+r^2+11 r \mu^2-4 \mu^4-5 \sqrt{r} \mu^3\right)\Big)\Big\}\nonumber\\
	&&+8 r^{3/2} \mu^2 \mathcal{A} \Big(6 \cos (2 \vartheta ) \left(\sqrt{r}-\mu\right)^2+\cos (4 \vartheta ) \left(5 \mu^2+2 \sqrt{r} \mu+r\right)-7 \mu^2+2 \sqrt{r} \mu-3 r\Big)\Big]\,,
	\end{eqnarray}
	where $\mu=\sqrt{r-2M}$. The obvious way how to solve this equation is to look for theories that satisfy $\tilde{f}_{TT}+2\tilde{f}_{TB}+\tilde{f}_{BB}=0$. However, having a closer look at this equation one finds that these theories are nothing but TEGR plus a cosmological constant since for Kerr we have $\lc{R}=0$ and then $T=B$. The non-trivial solution to this differential equation for $\mathcal{A}$ is
	\begin{eqnarray}\label{eq:H43b}
	\mathcal{A}(r,\vartheta)&=&\frac{\sin \vartheta  \cos ^2\vartheta  \left(4 \mu ^5+6 \mu ^2 r^{3/2}+r^{5/2}+4 \mu  r^2+\mu ^4 \sqrt{r}-16 \mu ^3 r\right)}{2 \mu ^2 r^{3/2} \left(-\mu ^2-4 \mu  \sqrt{r}+r\right)}+\mathcal{C}(r)F_1(r,\vartheta)+\mathcal{D}(r)F_2(r,\vartheta)\,,
	\end{eqnarray}
	where $\mathcal{C}(r)$ and $\mathcal{D}(r)$ are arbitrary functions (related to the integration of the differential equation) and $F_1(r,\vartheta)$ and $F_2(r,\vartheta)$ are specific functions which are related to the Legendre function of the first and second kind (see~\eqref{F2}-\eqref{F4}). Furthermore, by expanding the antisymmetric equation up to third order in $a$, one finds that $\mathcal{B}(r,\vartheta)=0$. 
	
	Thus fixing the tetrad component $H_{32}$ to be
	\begin{align}\label{kerrform}
	H_{32}(r,\vartheta)=   r \sin\vartheta+a^2\, \mathcal{A}(r,\vartheta)
	\end{align}
	and $\mathcal{A}$ as in \eqref{eq:H43b} we derived the good tetrad for a slowly rotating black hole spacetime in modified teleparallel gravity. 
	
	To display the torsion scalar and the boundary term, we choose the function $\mathcal{C}(r)=\mathcal{D}(r) = 0$. Up to third order in $a$ they become
	\begin{equation}
	\label{eq: Kerr a2 T}
	T=B=-\frac{2 \left(\sqrt{r}-\mu \right)^2}{\mu  r^{5/2}}-\frac{a^2 \left(2 \mu ^3+r^{3/2}+5 \mu ^2 \sqrt{r}+2 \mu  r\right) \left(\sqrt{r}-\mu \right)^2 \left(3 \mu ^2+\cos (2 \vartheta ) \left(5 \mu ^2+2 \mu  \sqrt{r}+r\right)+6 \mu  \sqrt{r}-r\right)}{2 \mu ^3 r^5 \left(-\mu ^2-4 \mu  \sqrt{r}+r\right)}\,.
	\end{equation}
	Since $-T+B=0$, we then have $\lc{R}=0$ as expected for the Kerr metric. One might also note, that since the torsion scalar \eqref{eq: Kerr a2 T} vanishes in the limit $r \rightarrow \infty$, the TEGR action integral constructed out of it could be considered as IR renormalised, like in Ref.\ \cite{Krssak:2015rqa}.
	
	\section{Solely axially symmetric branch: A first discussion}\label{sec5.solely}
	The physically most relevant axially symmetric tetrads so far are the ones we discussed in the previous section. Here we point to further classes of axially symmetric tetrads whose physical relevance still needs to be understood.
	
	\subsection{Stationary tetrads}\label{sec:2ndbranch}
	The discovery of the second branch of axially symmetric teleparallel Weitzenb\"ock tetrads is very recent and their physical relevance is not yet understood. Here we present a first discussion. A thorough investigation is left for ongoing and future research projects.
	
	
	For this branch and taking the tetrad~\eqref{tetradBrachB} we also find that there is only one remaining antisymmetric equation $E_{[23]}$, and it reads as follows
	\begin{eqnarray}
	0
	=&-\displaystyle\frac{1}{2\tilde{Q}(r,\vartheta)}\Big[
	(f_{T,\vartheta}+f_{B,\vartheta})\displaystyle\frac{\partial \tilde{Q}(r,\vartheta)}{\partial r}-(f_{T,r}+f_{B,r})\displaystyle\frac{\partial \tilde{Q}(r,\vartheta)}{\partial \vartheta}\Big]\,,\label{antibranch2}
	\end{eqnarray}
	where 
	\begin{equation}
	\tilde{Q}(r,\vartheta)=H_{23}(r,\vartheta)H_{00}(r,\vartheta)-H_{20}(r,\vartheta)H_{03}(r,\vartheta)\,.
	\end{equation}
	This equation is similar to the antisymmetric equation found for the regular branch (see~\eqref{antiGG}), just that $Q_r$ and $Q_\vartheta$ are generated by derivatives of the function $\tilde{Q}(r,\vartheta)$.
	
	Now we consider two cases. First, the easiest way to solve this equation is by imposing that $\tilde{Q}(r,\vartheta)=\textrm{const.}:=K$ yielding
	\begin{eqnarray}
	H_{23}(r,\vartheta)&=&\frac{H_{03}(r,\vartheta)H_{20}(r,\vartheta)}{H_{00}(r,\vartheta)}+\frac{K}{H_{00}(r,\vartheta)}\,.
	\end{eqnarray}
	The above form of $H_{23}(r,\vartheta)$ gives us the following form of the tetrad
	\begin{eqnarray}\label{goodtetrad44}
	h^A{}_\mu=\left(
	\begin{array}{cccc}
	H_{00}(r,\vartheta) & 0 & 0 & H_{03}(r,\vartheta) \\
	0 & H_{11}(r,\vartheta) & H_{12}(r,\vartheta) & 0 \\
	H_{20}(r,\vartheta) & 0 & 0 & \frac{H_{03}(r,\vartheta) H_{20}(r,\vartheta)+K}{H_{00}(r,\vartheta)} \\
	0 & -\frac{H_{11}(r,\vartheta) H_{12}(r,\vartheta)}{H_{32}(r,\vartheta)} & H_{32}(r,\vartheta) & 0 \\
	\end{array}
	\right)\,,
	\end{eqnarray}
	which solves the antisymmetric field equations for $f(T,B,\phi,X)$ without imposing any restriction in the form of $f$ nor the scalar field $\phi$. This tetrad gives us the following form of the metric
	\begin{eqnarray}
	ds^2&=&dt^2 \left(H_{00}(r,\vartheta)^2- H_{20}(r,\vartheta)^2\right)- H_{11}^2\Big(1+\frac{H_{12}(r,\vartheta)^2}{H_{32}(r,\vartheta)^2}\Big)dr^2- (H_{12}(r,\vartheta)^2+H_{32}(r,\vartheta)^2)d\vartheta^2\nonumber\\
	&& - \Big(\frac{(H_{03}(r,\vartheta) H_{20}(r,\vartheta)+K)^2}{H_{00}(r,\vartheta)^2}-H_{03}(r,\vartheta)^2\Big)d\varphi^2\nonumber\\
	&&+2\Big(H_{00}(r,\vartheta) H_{03}(r,\vartheta)-\frac{H_{20}(r,\vartheta) (H_{03}(r,\vartheta) H_{20}(r,\vartheta)+K)}{H_{00}(r,\vartheta)}\Big)dt \, d\varphi\,,
	\end{eqnarray}
	which does not contain any of the famous particular cases of the Pleba\'{n}ski–Demia\'{n}ski metric \eqref{PebDem} (not even the Schwarzschild metric). Moreover, when one tries to remove the cross term $dr\, d\varphi$ one finds that $g_{00}=-K^2/g_{33}$, which gives a metric that strongly deviates from the standard Schwarzschild-like form.
	
	The second case for which $\tilde{Q}\neq K$ can contain the Kerr metric as a special case, and similarly as it happens in the regular branch (previous section), one would need to replace the form of the scalar torsion and the boundary term in~\eqref{antibranch2} and then solve this equation for one of the functions $H_{ij}$. This procedure is again very involved in general and even for the Kerr metric, it is not easy to find a good tetrad. One finds that for the Schwarzschild case with their corresponding $H_{ij}$ (which are different to the $H_{ij}$ in~\eqref{SChwA}-\eqref{SChwB}), the form of $H_{32}$ cannot be $r\sin\vartheta$ to solve the antisymmetric equation (as in the principal branch). 
	Nevertheless, for the Kerr case with $M=0$ and also for the Minkowski case, if $H_{32}=r\sin\vartheta$, $T=B=0$ and then this choice solves the antisymmetric equation~\eqref{antibranch2}. Since this branch cannot have spherical symmetry, we will not study it further. 
	
	However, in the next section, we will show an explicit example of this situation with a good tetrad found previously in~\cite{Bejarano:2014bca}.
	
	\subsection{A time-dependent Kerr tetrad}
	
	Ref.~\cite{Bejarano:2014bca} gives a null tetrad that reproduces the Kerr metric and solves the antisymmetric equations with vanishing spin connection by having $T=0$. It can be translated into a regular tetrad in the Boyer-Lindquist coordinates as
	\begin{equation}
	h^A{}_\mu =
	\left(
	\begin{array}{cccc}
	\frac{e^{\lambda} (\Sigma - 2 M r) + e^{-\lambda} \Sigma }{2 \Sigma} &
	\frac{e^{\lambda} (\Sigma - 2 M r) - e^{-\lambda} \Sigma}{2 \Delta} &
	0 &
	\frac{a \sin^2 \vartheta \left( e^{\lambda} (\Sigma + 2 M r) - e^{-\lambda} \Sigma \right)}{2 \Sigma} \\
	\frac{e^{\lambda} (\Sigma - 2 M r) - e^{-\lambda} \Sigma }{2 \Sigma} & 
	\frac{e^{\lambda} (\Sigma - 2 M r) + e^{-\lambda} \Sigma}{2 \Delta} & 
	0 &
	\frac{a \sin^2 \vartheta \left( e^\lambda (\Sigma + 2 M r) + e^{-\lambda} \Sigma \right)}{2 \Sigma} \\
	0 &
	\frac{a^2 \sin \vartheta \, \cos \vartheta}{\Delta} &
	r &
	- a \sin \vartheta \, \cos \vartheta \\
	0 &
	-\frac{r a \sin \vartheta}{\Delta} &
	a \cos \vartheta &
	r \sin \vartheta 
	\end{array}
	\right) \,,
	\label{eq: Bejarano tetrad}
	\end{equation}
	where $a$ is the angular momentum parameter and $\Sigma$, $\Delta$ were defined in Eq.~\eqref{eq:Sigma Delta def}. One can check that it indeed reproduces the Kerr metric. It gives $T=0$ for vanishing spin connection when one introduces the time dependent function
	\begin{equation}\label{lambda}
	\lambda(t,r,\vartheta) = \frac{t(r^4 - 3 a^4 \cos^4 \vartheta -2 r^2 a^2 \cos^2 \vartheta - 4 a^2 M r \cos^2 \vartheta)}{2r (r^4 + a^4 \cos^4 \vartheta + 2 r^2 a^2 \cos^2 \vartheta)} + \tilde\lambda(r,\vartheta) \,,
	\end{equation}
	where $\tilde\lambda(r,\vartheta)$ is some arbitrary function. This then implies that antisymmetric equations are also solved.
	
	However, the caveat with this tetrad is that it does not satisfy the axial symmetry condition, since \eqref{eq: Bejarano tetrad} does not have the required $\cos \varphi$, $\sin \varphi$ dependence to belong to the first (regular) branch \eqref{eq: general axial tetrad}, nor does it depend only on $r$ and $\vartheta$ to belong to the second branch, solely axial \eqref{eq:2ndBranch}, see also Ref.~\cite{Hohmann:2019nat}.
	
	One easy way to understand the loss of spherical symmetry (in the teleparallel point of view) can be seen by considering the Schwarzschild  case ($a=0$)  in the tetrad~\eqref{eq: Bejarano tetrad} with \eqref{lambda}. In this case, $T=0$ and then the $f(T,B,\phi,X)$ antisymmetric equations are trivially satisfied (if $\phi=\phi(r)$). One might think that since $T=0$, the teleparallel quantities constructed from the torsion tensor do respect spherical symmetry. However, this is not the case since the torsion tensor presents non-zero components such as $T^{\varphi}{}_{\vartheta \varphi}=\cot\vartheta$ which violate the symmetry condition $\mathcal{L}_{X_\zeta}\Gamma^\sigma{}_{\mu\nu}=0$, see \eqref{eq:torsym}. Moreover, one can also compute other scalars using the irreducible pieces of the torsion tensor such as the $T_{\rm vec}, T_{\rm ax}$ or $T_{\rm ten}$~\cite{Hayashi:1979qx}, and notice that they depend on $r,$ $\vartheta$ and $t$ for the Schwarzschild case, which is indeed telling us that these scalars do not respect spherical symmetry (also they are not stationary). Moreover, if we consider the theory $f(T_{\rm vec},T_{\rm ax},T_{\rm ten})$~\cite{Bahamonde:2017wwk}, then one finds that the tetrad~\eqref{eq: Bejarano tetrad} does not solve its antisymmetric field equations and for the Schwarzschild case, they do depend on $\vartheta$ and $t$.
	
	\section{Determining inertial spin connections and Weitzenb\"ock tetrads by "switching off gravity"}\label{ssec:determinig_connection}
	Advancing from earlier ideas \cite{Krssak:2015rqa,Krssak:2018ywd} the authors of Ref.~\cite{Emtsova:2019moq} propose an algorithm 
	how to find the inertial spin connection to a given tetrad, without involving any field equations. This is an outstanding issue in TEGR where the antisymmetric field equations are identically satisfied. For modified teleparallel theories of gravity, this method gives tetrad-spin connection pairs (or Weitzenb\"ock tetrads) which solve the antisymmetric field equations for the spherical as well as spatially homogeneous and isotropic cases.
	
	We investigate the outcome of the algorithm in axial symmetry. We find that in a less symmetric situation the outcome of the algorithm is non-unique and does not necessarily produce a Weitzenb\"ock tetrad which solves the antisymmetric field equations of modified teleparallel theories of gravity.
	
	The suggested method to associate a spin connection to a tetrad is the following:
	\begin{enumerate}
		\item Consider a spacetime equipped with a metric $g$ and choose an arbitrary tetrad $h^A{}_\mu$.  Calculate the Levi-Civita spin connetion $\lc{\omega}^C{}_{D\mu} =h^C{}_\sigma h_D{}^\nu\lc{\Gamma}^\sigma{}_{\mu\nu}  - h_D{}^\nu \partial_\mu h^C{}_\nu$ and its corresponding Riemannian curvature tensor $\lc{R}^C{}_{D\mu\nu} = h^C{}_\sigma h_D{}^\rho \lc{ R}^\sigma{}_{\rho\mu\nu}$ in the frame basis, as displayed, where $\lc{\Gamma}^\sigma{}_{\mu\nu}$ are the Christoffel symbols of the metric and $\lc{R}^{\sigma}{}_{\rho\mu\nu}$ are the components of the Riemann curvature tensor derived with respect to the Levi-Civita connection in coordinate basis.
		\item Find a constraint either on the metric components or the coefficients appearing in the metric components, such that the Riemann curvature tensor vanishes, i.e. $ \lc{R}^\alpha{}_{\beta\mu\nu}=0$. This can be thought of as a limit where gravity is ``turned off''.
		\item Then, in general non-vanishing Levi-Civita connection of the metric evaluated on the flatness constraint $\lc{\omega}^C{}_{D\mu}|_{(\textrm{flatness condition})}$, represents a purely inertial flat spin connection, which is now identified with the teleparallel spin connection associated to the tetrad $h^A{}_\mu$.
		\item To obtain a Weitzenb\"ock gauge tetrad $h^A_W{}_\mu$ one searches the local Lorentz transformation $\Lambda^C{}_D(x)$ such that $\Lambda^C{}_B\partial_\mu (\Lambda^{-1})^B{}_D = \lc{\omega}^C{}_{D\mu}|_{(\textrm{flatness condition})}$.
	\end{enumerate}
	
	For Schwarzschild geometry, defined by the metric
	\begin{align}
	ds^2 = \left(1-\frac{r_s}{r}\right)dt^2 - \left(1-\frac{r_s}{r}\right) dr^2 - r^2 (d\vartheta^2 + \sin^2\vartheta d\varphi^2)
	\end{align}
	the Riemann curvature tensor depends on the Schwarzschild radius $r_s=2M$ and vanishes for $r_s = 0$. Using this condition, the above algorithm yields the off diagonal spherically symmetric Weitzenb\"ock tetrad (see Eq.~\eqref{tetrad} with \eqref{SChwA}-\eqref{SChwB}), that solves the antisymmetric field equations in modified teleparallel $f(T,B,\phi,X)$ theories of gravity.
	
	For homogeneous and isotropic FLRW geometry, defined by the metric
	\begin{align}
	ds^2 =  dt^2 - a(t)^2\left(\frac{dr^2}{1 - k r^2} + r^2 (d\vartheta^2+\sin^2\vartheta d\varphi^2)\right)\,,
	\end{align}
	the curvature tensor depends on the scale factor $a(t)$ and the spatial curvature parameter $k$. The condition $\dot a(t) + k = 0$ make the curvature tensor vanish and the above procedure yields the tetrad that has been found in the literature which solves antisymmetric field equations in general modified teleparallel theories of gravity~\cite{Hohmann:2019nat}.
	
	We now apply this algorithm, which was successful in highly symmetric situations, to Kerr, Taub-NUT, and C-metric geometries.
	\begin{enumerate}
		\item We choose a simple frame
		\begin{align}\label{eq:diag}
		h^A{}_\mu =
		\left(
		\begin{array}{cccc}
		\sqrt{g_{00}} & 0 & 0 & \frac{g_{03}}{\sqrt{g_{00}}} \\
		0 & \sqrt{g_{11}}  & 0  & 0 \\
		0 & 0 & \sqrt{g_{22}}  & 0 \\
		0 & 0 & 0 & \sqrt{\frac{g_{03}^2}{g_{00}} + g_{33}} \\
		\end{array}
		\right)\,,
		\end{align}
		of the line element $ds^2= g_{00} dt^2 + g_{03} dt \,d\varphi + g_{11} dr^2 + g_{22} d\vartheta^2 + g_{33} d\varphi^2$, where all metric functions only depend on $r$ and $\vartheta$, $g_{\mu\nu} = g_{\mu\nu}(r,\vartheta)$. This metric includes the whole Pleba\'nski-Demia\'nski class we introduced in~\eqref{PebDem} and thus in particular the famous special cases:
		\begin{itemize}
			\item The Kerr metric for the choice 
			\begin{align}
			g_{00} &= 1 -  \frac{2 M r}{\Sigma}\,,\quad g_{03} =  \frac{4 M r a \sin^2\vartheta}{\Sigma}\,,\quad g_{11} = -\frac{\Sigma}{\Delta},\quad  g_{22} = -\Sigma\,,\nonumber\\
			g_{33} &=-\Big(r^2 + a^2 + \frac{ 2 M r  a^2 \sin^2\vartheta}{\Sigma}\Big)\sin^2\vartheta\,,\nonumber
			\end{align}
			where $\Sigma$ and $\Delta$ are defined below \eqref{kerrC}.
			\item The Taub-NUT metric for
			\begin{align}
			g_{00} &= 1 - \frac{2(b^2 + M r)}{b^2+r^2}\,,\quad g_{03} = 2b (\chi - \cos\vartheta)\left(1 - \frac{2(b^2 + M r)}{b^2+r^2}\right)\,,\quad g_{11} = -\frac{b^2+r^2}{r^2-2Mr-b^2}\nonumber\\
			g_{22} &=-(b^2+r^2)\,,\quad g_{33} =  4b^2 (\chi - \cos\vartheta)\left(1 - \frac{2(b^2 + M r)}{b^2+r^2}\right) - (b^2 + r^2)\sin^2\vartheta\nonumber\,.
			\end{align}
			\item The C-metric for
			\begin{eqnarray}
			g_{00} = \frac{\mathcal{Q}}{r^2\Omega^2}\,,\quad g_{11} = -\frac{r^2}{\Omega^2 \mathcal{Q}}\,,\quad g_{22} = -\frac{r^2}{ \Omega^2\mathcal{P}}\,,\quad g_{33} = -\frac{\mathcal{P}}{\Omega^2}r^2\sin^2\vartheta\,,\nonumber
			\end{eqnarray}
			where $Q$, $\Omega$ and $P$ are defined in \eqref{CmetricFunc}.
		\end{itemize} 
		Calculating the frame components of the Riemann curvature tensor in the frame \eqref{eq:diag} results in curvature components depending on the parameters of the metric. All metrics share the parameter $M$ and in addition: the rotation $a$ for the Kerr metric, the NUT parameter $b$ for the Taub-NUT metric, and $\alpha$ for the C-metric.
		\item To switch gravity off, a condition on the parameters is searched such that the curvature tensor $\lc{R}^{\lambda}{}_{\mu\nu\rho}$ vanishes. It could be a relation of the type $M=M(Y)$, where $Y$ is one of the parameters of the metric in consideration, or finding a value which $Y$ and $M$ assumed. The conditions for the different cases are
		\begin{itemize}
			\item Kerr: $M=0$ suffices to make all components of the curvature tensor  vanish. It is optional to also set $a$ to a fixed value, for example $a=0$;
			\item Taub-NUT: $M=0$ and $b=0$ is necessary to make all components of the curvature tensor vanish;
			\item C-metric: $M=0$ and $\alpha=0$ is necessary to make all components of the curvature tensor vanish;
		\end{itemize}
		\item 
		For the different geometries listed before, $\lc{\omega}^C{}_{D\mu}|_{M=0}$,   $\lc{\omega}^C{}_{D\mu}|_{(M=0,b=0)}$ and $\lc{\omega}^C{}_{D\mu}|_{(M=0,\alpha=0)}$, respectively, represent flat spin connections. It turns out that these flat spin connections coincide for all three geometries under the conditions $M=a=0$ for Kerr, $M=b=0$ for Taub-NUT and $M=\alpha=0$ for the C-metric. 
		
		Starting with the Kerr geometry case, identifying the just determined flat spin connection with the teleparallel connection generated by local Lorentz transformations $\lc{\omega}^C{}_{D\mu}|_{M=0}=\Lambda^C{}_B\partial_\mu (\Lambda^{-1})^B{}_D$, yields an equation to determine $\Lambda^C{}_D = \Lambda^C{}_D(r,\vartheta,\varphi)$. Solving the equivalent equation $(\Lambda^{-1})^C{}_B \lc{\omega}^B{}_{D\mu}|_{M=0} = \partial_\mu (\Lambda^{-1})^C{}_D $ for Kerr geometry yields
		\begin{align}
		(\Lambda^{-1})^C{}_D =
		\left(
		\begin{array}{cccc}
		1 & 0 & 0 & 0 \\[\medskipamount]
		0 & \displaystyle\frac{r \cos \varphi \sin \vartheta}{\sqrt{\Sigma}}  & \sqrt{\frac{ 2(a^2 + r^2)}{a^2 + r^2 + \Sigma}} \cos\varphi \cos \vartheta  & - \sin\varphi \\[\medskipamount]
		0 & \displaystyle\frac{r \sin \varphi \sin \vartheta}{\sqrt{\Sigma}} & \sqrt{\frac{ 2(a^2 + r^2)}{a^2 + r^2 + \Sigma}} \sin\varphi \cos \vartheta  & \cos \varphi \\[\medskipamount]
		0 & \sqrt{\frac{a^2 + r^2}{\Sigma}}\cos\vartheta & -\sqrt{\frac{2}{a^2 + r^2 + \Sigma}}r \sin\vartheta & 0 \\[\medskipamount]
		\end{array}
		\right)\,.
		\end{align}
		Since the spin connections of the different geometries coincide when all parameters in the metrics vanish, we can simply set $a\to0$ to obtain the desired Lorentz transformation for the other two cases. This also means that for the Kerr case the algorithm yields two possible Lorentz transformations $\Lambda$ and $\Lambda|_{a=0}$.
		\item The Weitzenb\"ock tetrad then is obtained by applying the Lorentz transformation to the tetrad \eqref{eq:diag} $h^C_W{}_\mu = h^D{}_\mu (\Lambda^{-1})^C{}_D$. Most interestingly one obtains the reduced axially symmetric tetrads \eqref{tetrad} for all geometries with a fixed $H_{32}$ component.
		\begin{itemize}
			\item Kerr geometry: Using $(\Lambda^{-1})^A{}_B$ to generate the Weitzenb\"ock tetrad yields the tetrad components \eqref{kerrA}-\eqref{kerrB} with $H_{32} = r \sin\vartheta$. Using instead $(\Lambda^{-1})^C{}_D|_{a=0}$ yields the tetrad components \eqref{kerrA}-\eqref{kerrB} with $H_{32} = \sqrt{r^2 + a^2 \cos^2\vartheta}\sin\vartheta$. Comparing these tetrad components with the ones we found from solving the antisymmetric field equations in Sec.~\ref{sec:case6}, we conclude that neither of the derived tetrads is a solution.
			\item Taub-NUT geometry: In this case only one Lorentz transformation, $(\Lambda^{-1})^A{}_B|_{a=0}$, is available to generate the Weitzenb\"ock tetrad. It precisely becomes the reduced axially symmetric tetrad with components \eqref{TN} and $H_{32} = \sqrt{b^2 + r^2}\sin\vartheta$. Comparing this result with the solutions found in Sec.~\ref{sec:2} we see that this tetrad indeed is a solution of the antisymmetric field equations.
			\item The C-metric: As for the Taub-NUT case only one Lorentz transformation is available to generate the Weitzenb\"ock tetrad. This time we obtain the reduced axially symmetric tetrad with components \eqref{CM} and $H_{32} = \frac{r \sin\vartheta}{P \Omega}$. As in the Kerr case, this tetrad does not solve the antisymmetric field equations.
		\end{itemize}
	\end{enumerate}
	
	Thus the algorithm suggested in \cite{Emtsova:2019moq} associates spin connections to tetrads, respectively determines Weitzenb\"ock tetrads. The procedure is however non-unique, and, sadly, has no clear direct connection to finding solutions to the antisymmetric field equations of modified teleparallel theories of gravity. The algorithm may have relevance for TEGR, where the antisymmetric equations are identically satisfied. Whether there is any way a distinguished physical interpretation of the tetrads we found using the algorithm needs to be investigated.
	
	\section{Conclusions}\label{sec:conclusion}
	In the present work, we address the study of axially symmetric teleparallel geometries in the framework of $f(T,B,\phi,X)$ gravity. For this task, we emphasise the existence of two different branches of axially symmetric tetrad fields and Lorentz flat spin connections, i.e.\ they preserve the underlying symmetry conditions under the action of the rotation group $SO(2)$, as was shown in~\cite{Hohmann:2019nat} (see Sec.~\ref{sec:twobranches}). The important difference between the branches is the fact that the first, the regular branch (see Sec.~\ref{ssec:regbranch}) is consistent with a spherically symmetric teleparallel geometry in a certain limit, in the sense of teleparallel geometry, see \eqref{eq:tpsymm}, while the second, the solely axially symmetric branch (see Sec.~\ref{sec:alternative}) cannot have such a limit.
	
	In teleparallel theories of gravity, the field equations can be decomposed into symmetric and, in general nontrivial, antisymmetric parts. The first step towards a solution is always to solve the antisymmetric field equations, which is mostly done in the Weitenzb\"ock gauge where all degrees of freedom are encoded in the tetrad and the spin connection is set to zero. The tetrads found this way are often called good tetrads, which serve as an ansatz that is fed into the symmetric field equations.
	
	We focused on finding solutions to the antisymmetric field equations of the $f(T,B,\phi,X)$ class of gravity theories, which contains many modified teleparallel theories of gravity discussed in the literature, in axial symmetry, starting from the regular branch, before we discussed alternative tetrad choices.  In particular, we started the search for teleparallel generalisations of axially symmetric spacetimes beyond the Pleba\'nski-Demia\'nski class of solutions of general relativity.
	
	We introduce the first categorisation for teleparallel axially symmetric spacetimes and demonstrate the existence of a good tetrad containing the Taub-NUT subclass of the Pleba\'nski-Demia\'nski metric in the regular branch (i.e. allowing a continuous reduction to the spherically symmetric case). The physical viability of this geometry has recently revisited in virtue of the absence of pathologies for free-falling observers and thermodynamics \cite{Clement:2015cxa,Kubiznak:2019yiu,Bordo:2019tyh}. In addition, considering the highly nonlinear character of the field equations of $f(T,B,\phi,X)$, we are also able to obtain an analytical expression for the good tetrad in Kerr geometry under the slow rotation approximation, which constitutes a first preliminary and promising result for the achievement of a complete solution describing rotating black holes in modified teleparallel gravity. 
	
	The corresponding good tetrads related to the aforementioned configurations and other particular cases can be summarised as follows:
	\begin{enumerate}
		\item For the regular branch:
		\begin{enumerate}
			\item An axially symmetric universal good tetrad~\eqref{goodtetrad1} for any $f(T,B,\phi,X)$ which turns out not to reproduce any of the special cases of the vacuum Pleba\'nski-Demia\'nski metric (except the Schwarzschild case). This tetrad generalises the one found in~\cite{Jarv:2019ctf}.
			\item An axially symmetric good tetrad~\eqref{tetradgood} which is a generalisation of the standard spherically symmetric tetrad~\eqref{spherical} and reproduces a metric behaving like a family of Taub-NUT-like metrics~\eqref{metrictaublike}. This tetrad is valid for any $f(T,B,\phi,X)$ provided by a scalar field of the form $\phi=\phi(r)$.
			\item An axially symmetric good tetrad~\eqref{goodtetrad3} which shows a high dependence on the angular coordinate $\vartheta$, and provides the nontrivial relation $g_{00}=-\,g_{03}/2$ for the metric tensor. This tetrad is valid for any $f(T,B,\phi,X)$ with a scalar field being $\phi=\phi(\vartheta)$. As is shown, this result strongly constrains the form of the metric tensor, so a further study of the symmetric equations would be necessary in order to clarify whether this tetrad is physically viable or not.
			\item A good tetrad for a slowly rotating ($a\ll 1$) Kerr metric given by~\eqref{kerrA}-\eqref{kerrB} with~\eqref{kerrform}. This result is only valid for $f(T,B)$ gravity. Even though this good tetrad solves the antisymmetric field equations, all the metric (and tetrad) functions are determined. It can be used to find perturbative teleparallel modifications of slowly rotating axially symmetric solutions around Kerr (see~\cite{Konno:2009kg, Yunes:2009hc, Pani:2009wy, Pani:2011gy, Ayzenberg:2014aka,Maselli:2015tta,Cano:2019ore} for similar works in curvature-based theories of gravity) or other analyses where the Kerr metric is assumed.
			\item A new spherically symmetric good tetrad~\eqref{spherical2} for $f(T)$ gravity having $T=0$ whose metric is constraint to have a $g_{00}=-\,1/g_{11}$ form. This tetrad only obeys spherical symmetry in the trivial case when $g_{00}=1$, otherwise the teleparallel connection does not satisfy the symmetry condition, i.e.,  $L_\zeta \Gamma\neq0$.
		\end{enumerate}
		\item For the solely axially symmetric branch:
		\begin{enumerate}
			\item An axially symmetric good tetrad~\eqref{goodtetrad44} which does not have any Pleba\'nski-Demia\'nski metric as particular cases. Even the Schwarzschild metric cannot be recovered with this tetrad. Furthermore, the $t$-$t$ component is related to the $\varphi$-$\varphi$ component as $g_{00}=-\,K^2/g_{33}$, which gives us a metric behaving very differently to any standard known vacuum solution in GR. 
		\end{enumerate}
	\end{enumerate}
	
	In the study of extended gravitational theories beyond GR, the search and analysis of solutions to the underlying field equations are fundamental to figure out their dynamical properties and to test their validity in different astrophysical and cosmological situations. In this regard, the consideration of rotating black holes which may carry additional charges as is the case in axial symmetry turns out to be essential for a phenomenological assessment of such theories in terms of a realistic configuration. Black hole angular momentum measurements include observations on the dynamics of accretion disks and stellar objects in their vicinity, the study of shadow images, or the detection of gravitational waves \cite{Teukolsky:2014vca,Abbott:2016blz,Akiyama:2019cqa,Abbott:2016izl}. The possible effects of a gravitomagnetic monopole on the black hole shadow and on the twist of light in microlensing events have been considered for the design of future tests of axially symmetric configurations with a NUT charge \cite{LyndenBell:1996xj,Rahvar:2002es,Rahvar:2003fh,Abdujabbarov:2012bn}. In this sense, it is also expected to obtain new phenomenological constraints for the viability of generalised axially symmetric teleparallel geometries as the ones presented in this work. In addition, further extensions of slowly and rigidly rotating, stationary and axially symmetric bodies, e.g. see \cite{Hartle:1968si,Hartle:1967he}, may be considered by setting as a background spacetime the slowly rotating configuration found in this work, in order to include higher-order multipole moments in the gravitational scheme and analyse the effect of rotation on stellar structures in the realm of modified teleparallel gravity. Further research following these lines will be addressed in future works.

	\begin{acknowledgments}
		This research was supported by the European Regional Development Fund through the Center of Excellence TK133 ``The Dark Side of the Universe''. C. P. and L. J. were supported by the Estonian Ministry for Education and Science through the Estonian Research Council grants No. PSG489 and No. PRG356, respectively, while S.B. and J.G.V. were supported by the European Regional Development Fund via the Mobilitas Pluss programme grants MOBJD423 and MOBJD541, respectively.
	\end{acknowledgments}
	
	
	\appendix
	

	\section{Torsion Scalar and Boundary term in axial symmetry - Principal branch}\label{ssec:torsion_scalar}
	The torsion scalar and the boundary term for the general tetrad~\eqref{tetrad} in axial symmetry become
	\begin{eqnarray}
	T&=&\frac{1}{2 H_{00}^2 H_{11}^2 H_{23}^2 H_{32} \left(H_{12}^2+H_{32}^2\right)^2}\Big[\Big(\left(H_{12}^2 H_{03,r}^2+4 H_{12} H_{23} H_{12,r} H_{23,r}+H_{32} \left(H_{32} H_{03,r}^2+4 H_{23} H_{23,r} H_{32,r}\right)\right) H_{32}^3\nonumber\\
	&&-4 H_{11} H_{23} \Big(H_{11,\vartheta} H_{12}^3+\left(H_{32} H_{32,r}-H_{11,\vartheta} H_{23,\vartheta}\right) H_{12}^2+H_{32} \left(H_{32} H_{11,\vartheta}+H_{32,\vartheta} H_{23,r}-H_{23,\vartheta} H_{32,r}\right) H_{12}\nonumber\\
	&&+H_{32}^2 \left(-H_{11,\vartheta} H_{23,\vartheta}+H_{12,r} H_{23,\vartheta}-H_{12,\vartheta} H_{23,r}+H_{32} H_{32,r}\right)\Big) H_{32}+H_{11}^2 \Big(4 H_{23} H_{32,\vartheta} H_{12}^3+\nonumber\\
	&&\left(H_{32} \left(H_{03,\vartheta}^2-4 H_{23} H_{12,\vartheta}\right)-4 H_{23} H_{23,\vartheta} H_{32,\vartheta}\right) H_{12}^2+4 H_{23} H_{32} \left(H_{12,\vartheta} H_{23,\vartheta}+H_{32} H_{32,\vartheta}\right) H_{12}\nonumber\\
	&&+H_{32}^3 \left(H_{03,\vartheta}^2-4 H_{23} H_{12,\vartheta}\right)\Big)\Big) H_{00}^2-2 \Big(H_{00,r} \Big(H_{03} \left(H_{12}^2+H_{32}^2\right) H_{03,r}-2 H_{23} \Big(H_{23,r} H_{12}^2+H_{23} H_{12,r} H_{12}\nonumber\\
	&&+H_{32}^2 H_{23,r}+H_{23} H_{32} H_{32,r}\Big)\Big) H_{32}^3-2 H_{11} H_{23} \Big(H_{23} \left(H_{12,\vartheta} H_{00,r}+H_{00,\vartheta} \left(H_{11,\vartheta}-H_{12,r}\right)\right) H_{32}^2\nonumber\\
	&&+H_{12} H_{23} \left(H_{00,\vartheta} H_{32,r}-H_{32,\vartheta} H_{00,r}\right) H_{32}-H_{32}^4 H_{00,r}+H_{12}^2 \left(H_{23} H_{00,\vartheta} H_{11,\vartheta}-H_{32}^2 H_{00,r}\right)\Big) H_{32}\nonumber\\
	&&+H_{11}^2 H_{00,\vartheta} \Big(2 H_{23} H_{32} H_{12}^3+\left(2 H_{32,\vartheta} H_{23}^2-2 H_{32} H_{23,\vartheta} H_{23}+H_{03} H_{32} H_{03,\vartheta}\right) H_{12}^2\nonumber\\
	&&+2 H_{23} H_{32} \left(H_{32}^2-H_{23} H_{12,\vartheta}\right) H_{12}+H_{32}^3 \left(H_{03} H_{03,\vartheta}-2 H_{23} H_{23,\vartheta}\right)\Big)\Big) H_{00}\nonumber\\
	&&+H_{03}^2 H_{32} \left(H_{12}^2+H_{32}^2\right) \left(H_{11}^2 H_{00,\vartheta}^2+H_{32}^2 H_{00,r}^2\right)\Big]\,,\label{Tgeneral}
	\end{eqnarray}
	\begin{eqnarray}
	B&=&\frac{1}{H_{00} H_{11}^3 H_{23} H_{32}^2 (H_{12}^2+H_{32}^2)^3}\Big[2 \Big\{-(H_{12}^2+H_{32}^2) H_{11,r} \Big((H_{23} H_{00,r}+H_{00} H_{23,r}) H_{12}^2+H_{00} H_{23} H_{12,r} H_{12}\nonumber\\
	&&+H_{32} (H_{00} H_{32} H_{23,r}+H_{23} (H_{32} H_{00,r}+H_{00} H_{32,r}))\Big) H_{32}^4+H_{11} \Big(\Big((H_{23} H_{00,r}+H_{00} H_{23,r}) H_{32,r}\nonumber\\
	&&+H_{32} \left(2 H_{00,r} H_{23,r}+H_{23} H_{00,rr}+H_{00} H_{23,rr}\right)\Big) H_{12}^4+\Big(H_{00} H_{32} H_{12,r} H_{23,r}+H_{23} (H_{00} H_{12,r} H_{32,r}\nonumber\\
	&&+H_{32} \left(H_{00,r} H_{12,r}+H_{00} H_{12,rr}\right))\Big) H_{12}^3+H_{32} \Big(H_{32} (3 H_{23} H_{00,r} H_{32,r}+2 H_{32} \left(2 H_{00,r} H_{23,r}+H_{23} H_{00,rr}\right))\nonumber\\
	&&+H_{00} \left(H_{32} \left(3 H_{23,r} H_{32,r}+2 H_{32} H_{23,rr}\right)+H_{23} \left(-H_{12,r}^2+2 H_{32,r}^2+H_{32} H_{32,rr}\right)\right)\Big) H_{12}^2\nonumber\\
	&&+H_{32}^2 \left(H_{00} H_{32} H_{12,r} H_{23,r}+H_{23} \left(H_{32} \left(H_{00,r} H_{12,r}+H_{00} H_{12,rr}\right)-3 H_{00} H_{12,r} H_{32,r}\right)\right) H_{12}\nonumber\\
	&&+H_{32}^3 (H_{32} \left(2 H_{23} H_{00,r} H_{32,r}+H_{32} \left(2 H_{00,r} H_{23,r}+H_{23} H_{00,rr}\right)\right)+H_{00} (H_{32} \left(2 H_{23,r} H_{32,r}+H_{32} H_{23,rr}\right)\nonumber\\
	&&+H_{23} \left(H_{12,r}^2+H_{32} H_{32,rr}\right)))\Big) H_{32}^3-H_{11}^2 \left(H_{12}^2+H_{32}^2\right) (H_{32} (H_{00,r} H_{32}^4+H_{23} (H_{00,\vartheta} \left(H_{12,r}-2 H_{11,\vartheta}\right)\nonumber\\
	&&-H_{12,\vartheta} H_{00,r}) H_{32}^2+H_{12} H_{23} \left(H_{32,\vartheta} H_{00,r}-H_{00,\vartheta} H_{32,r}\right) H_{32}+H_{12}^2 \left(H_{32}^2 H_{00,r}-2 H_{23} H_{00,\vartheta} H_{11,\vartheta}\right))\nonumber\\
	&&+H_{00} (H_{32} H_{11,\vartheta} H_{12}^3+\left(H_{32,r} H_{32}^2-\left(2 H_{11,\vartheta} H_{23,\vartheta}+H_{23} H_{11,\vartheta\vartheta}\right) H_{32}+2 H_{23} H_{11,\vartheta} H_{32,\vartheta}\right) H_{12}^2\nonumber\\
	&&+H_{32} \left(H_{11,\vartheta} H_{32}^2+\left(H_{32,\vartheta} H_{23,r}-H_{23,\vartheta} H_{32,r}\right) H_{32}-H_{23} H_{11,\vartheta} H_{12,\vartheta}\right) H_{12}\nonumber\\
	&&+H_{32}^2 \left(H_{32,r} H_{32}^2-\left(2 H_{11,\vartheta} H_{23,\vartheta}-H_{12,r} H_{23,\vartheta}+H_{23} H_{11,\vartheta\vartheta}+H_{12,\vartheta} H_{23,r}\right) H_{32}+H_{23} H_{11,\vartheta} H_{32,\vartheta}\right))) H_{32}\nonumber\\
	&&+H_{11}^3 \Big[H_{32} \left(H_{00} H_{32,\vartheta}-H_{32} H_{00,\vartheta}\right) H_{12}^5+(H_{32} \left(H_{32} \left(2 H_{00,\vartheta} H_{23,\vartheta}+H_{23} H_{00,\vartheta\vartheta}\right)-2 H_{23} H_{00,\vartheta} H_{32,\vartheta}\right)\nonumber\\
	&&-H_{00} \left(\left(H_{12,\vartheta}-H_{23,\vartheta\vartheta}\right) H_{32}^2+\left(2 H_{23,\vartheta} H_{32,\vartheta}+H_{23} H_{32,\vartheta\vartheta}\right) H_{32}-2 H_{23} H_{32,\vartheta}^2\right)) H_{12}^4\nonumber\\
	&&+H_{32} (-2 H_{00,\vartheta} H_{32}^3+2 H_{00} H_{32,\vartheta} H_{32}^2+\left(H_{00} H_{12,\vartheta} H_{23,\vartheta}+H_{23} \left(H_{00,\vartheta} H_{12,\vartheta}+H_{00} H_{12,\vartheta\vartheta}\right)\right) H_{32}\nonumber\\
	&&-H_{00} H_{23} H_{12,\vartheta} H_{32,\vartheta}) H_{12}^3-H_{32}^2 (H_{32} \left(3 H_{23} H_{00,\vartheta} H_{32,\vartheta}-2 H_{32} \left(2 H_{00,\vartheta} H_{23,\vartheta}+H_{23} H_{00,\vartheta\vartheta}\right)\right)\nonumber\\
	&&+H_{00} \left(2 \left(H_{12,\vartheta}-H_{23,\vartheta\vartheta}\right) H_{32}^2+\left(3 H_{23,\vartheta} H_{32,\vartheta}+H_{23} H_{32,\vartheta\vartheta}\right) H_{32}+H_{23} \left(H_{12,\vartheta}^2-4 H_{32,\vartheta}^2\right)\right)) H_{12}^2\nonumber\\
	&&+H_{32}^3 (H_{00} H_{32,\vartheta} H_{32}^2+\left(H_{00} H_{12,\vartheta} H_{23,\vartheta}+H_{23} \left(H_{00,\vartheta} H_{12,\vartheta}+H_{00} H_{12,\vartheta\vartheta}\right)\right) H_{32}-H_{32}^3 H_{00,\vartheta}\nonumber\\
	&&-5 H_{00} H_{23} H_{12,\vartheta} H_{32,\vartheta}) H_{12}+H_{32}^4 (H_{32} \left(H_{32} \left(2 H_{00,\vartheta} H_{23,\vartheta}+H_{23} H_{00,\vartheta\vartheta}\right)-H_{23} H_{00,\vartheta} H_{32,\vartheta}\right)\nonumber\\
	&&+H_{00} \left(\left(H_{23,\vartheta\vartheta}-H_{12,\vartheta}\right) H_{32}^2-H_{23,\vartheta} H_{32,\vartheta} H_{32}+H_{23} H_{12,\vartheta}^2\right))\Big]\Big\}\Big]\,.\label{Bgeneral}
	\end{eqnarray}
	
	\section{Kerr perturbations}\label{ssec:kerr_perturbations}
	The form of the functions appearing in~\eqref{eq:H43b} are
	\begin{eqnarray}
	F_2(r,\xi)&=&\sqrt{\xi^2-1}\exp\Big[\int (\xi-\xi^3)^{-1}d\xi\Big]P(f_2(r),1,\xi)\,,\label{F2}\\
	F_4(r,\xi)&=&\sqrt{\xi^2-1}\exp\Big[\int (\xi-\xi^3)^{-1}d\xi\Big]Q(f_2(r),1,\xi)\,,\label{F4}\\
	f_2(r)&=&\frac{1}{2M}\Big[\sqrt{(32r^{3/2}-48M\sqrt{r})\sqrt{r-2M}+41M^2-80Mr+32r^2}-M\Big]\,,
	\end{eqnarray}
	where $\xi=\cos\vartheta$ and $P(x,y,z)$ and $Q(x,y,z)$ are the Legendre functions of the first and second kind respectively~\cite{whittaker2020course}.
	
	
	\bibliographystyle{utphys}
	\bibliography{AxSymGT}
	
\end{document}